\documentclass[fleqn,usenatbib]{mnras}

\usepackage[T1]{fontenc}

\DeclareRobustCommand{\VAN}[3]{#2}
\let\VANthebibliography\thebibliography
\def\thebibliography{\DeclareRobustCommand{\VAN}[3]{##3}\VANthebibliography}

\usepackage{graphicx}	
\usepackage{amsmath}	
\usepackage{amssymb}	
\usepackage[export]{adjustbox}

\usepackage{morefloats}
\usepackage{newtxtext,newtxmath}
\usepackage[final]{pdfpages}
\usepackage{subcaption}

\def\ha{H$\alpha$}
\def\hb{H$\beta$}

\def\kms{$\rm km\,s^{-1}$}
\def\str{{\sc starlight}}
 
\def\st{{\sc starlight}}
\newcommand{\rev}[1]{{\textcolor{black}{#1}}}

\newcommand\footnoteref[1]{\protected@xdef\@thefnmark{\ref{#1}}\@footnotemark}

\title[AGNs stellar population and excitation through {\sc megacubes}]{Mapping the stellar population and gas excitation of MaNGA galaxies with {\sc megacubes}. Results for AGN versus control sample}
\author[Riffel, R. et al.]{Rog\'erio Riffel$^{1,2,4}$\thanks{E-mail: riffel@ufrgs.br},
 Nicolas D. Mallmann$^{1,2}$, 
 Sandro B. Rembold$^{3,2}$,
 Gabriele S. Ilha$^{3,2,7}$,
 \newauthor
 Rogemar A. Riffel$^{3,2}$,
 Thaisa Storchi-Bergmann$^{1,2}$,
 Daniel Ruschel-Dutra$^{5}$,
 Alexandre Vazdekis $^{4,6}$,
 \newauthor
 Ignacio Mart\'\i n-Navarro$^{4,6}$,
 Jaderson S. Schimoia$^{3,2}$,
 Cristina Ramos Almeida$^{4,6}$,
 Luiz N. da Costa$^{2}$,
  \newauthor
 Glauber C. Vila-Verde$^{2}$,
 Lara Gatto$^{1,2}$
\\
$^{1}$ Departamento de Astronomia, Instituto de F\'\i sica, Universidade Federal do Rio Grande do Sul, CP 15051, 91501-970, Porto Alegre, RS, Brazil \\
$^{2}$Laborat\'orio Interinstitucional de e-Astronomia - LIneA, Rua Gal. Jos\'e Cristino 77, Rio de Janeiro, RJ - 20921-400, Brazil\\
$^{3}$ Departamento de F\'\i sica, Centro de Ci\^encias Naturais e Exatas, Universidade Federal de Santa Maria, 97105-900, Santa Maria, RS, Brazil \\ 
$^{4}$ Instituto de Astrof\'\i sica de Canarias, Calle V\'\i a L\'actea s/n, E-38205 La Laguna, Tenerife, Spain\\
$^5$Departamento de F\'isica, Universidade Federal de Santa Catarina, P.O. Box 476, 88040-900, Florian\'opolis, SC, Brazil\\
$^6$ Departamento de  Astrof\'\i sica, Universidad de La Laguna, E-38205, Tenerife, Spain
$^7$ Universidade do Vale do Para\'iba. Av. Shishima Hifumi, 2911, CEP: 12244-000, São Jos\'e dos Campos, SP, Brazil\\
}

\date{Accepted XXX. Received YYY; in original form ZZZ}

\pubyear{2023}

\begin{document}
\label{firstpage}
\pagerange{\pageref{firstpage}--\pageref{lastpage}}
\maketitle

\begin{abstract}
We present spaxel-by-spaxel stellar population fits for the $\sim$10 thousand MaNGA datacubes. We provide multiple extension fits files, nominated as {\sc megacubes}, with maps of several properties as well as emission-line profiles that are provided for each spaxel. All the {\sc megacubes} are available through a web interface (\url{https://manga.linea.org.br/} or \url{http://www.if.ufrgs.br/~riffel/software/megacubes/}). We also defined a final Active Galactic Nuclei (AGN) sample, as well as a control sample matching the AGN host galaxy properties. 
We have analysed the stellar populations and  spatially resolved emission-line diagnostic diagrams of these AGNs and compared them with the control galaxies sample.  We find that the relative fractions of young ($t \leq $56\,Myr) and intermediate-age  (100\,Myr~$\leq t \leq$~2\,Gyr) show predominantly a positive gradient for both AGNs and controls. The relative fraction of intermediate-age stellar population is higher in AGN hosts when compared to the control sample, and this difference becomes larger for higher [\ion{O}{iii}] luminosity AGNs. We attribute this to the fact that extra gas is available in these more luminous sources and that it most likely originates from mass-loss from the intermediate-age stars. The spatially resolved diagnostic diagrams reveal that the AGN emission is concentrated in the inner 0.5\,$R_e$ (effective radius) region of the galaxies, showing that the AGN classification is aperture dependent
and that emission-line ratios have to be taken together with the \ha\ equivalent width for proper activity classification. We present a composite ``BPT+WHAN" diagram that produces a more comprehensive mapping of the gas excitation.

\end{abstract}




\section{Introduction}

The evolution of the present-day galaxies during cosmic time is the result of the transformation of molecular gas into stars, the production (and release) of metals by stellar evolution, and the interaction among these processes and their environments (e.g. internal and external processes). All these evolution/transformation processes leave signatures in the observed properties of
 galaxies that we can analyze and thus reconstruct the galaxies' star formation histories (SFHs). The study of this fossil record is key to our understanding of the structure, composition, and evolution of galaxies \citep[e.g.][and references therein]{CidFernandes+04,CidFernandes+05,Conselice+14,Tacconi+20,Sanchez+20,Sanchez+21}.

 Significant improvement in the understanding of galaxy evolution processes has been reached in the last decades, primarily due to new results based on large spectroscopic and imaging surveys such as the Sloan Digital Sky Survey \citep[SDSS,][]{York+00}. These studies have shown that galaxies 
 can roughly be divided into passive and star-forming. The passive galaxies are not forming (or forming at a very low rate) stars and host a red and old stellar population (e.g. the red-and-dead galaxies), while the star-forming galaxies are blue, hosting large fractions of young stellar populations. Such bi-modal behaviour has been observed  in a number of studies over the years  \citep[e.g.][]{Baldry+04,Wetzel+12,Kauffmann+03a,Noeske+07,Wel+14,Muzzin+13,Brammer+09}. However, so far it is not yet clear which mechanisms are stopping the star formation process and transforming the blue star-forming spiral galaxies into {\it red-and-dead} galaxies, being thus a major challenge in modern astrophysics to understand the nature of the physical mechanisms quenching star formation in galaxies.

One mechanism that a number of studies have invoked to explain the suppression of the star-formation in galaxies is the active galactic nuclei (AGN) feedback. It can quench star formation by (re)moving and/or heating the gas. In this context, AGN outflows are considered a process that suppresses star-formation \citep[e.g. as negative feedback, see][and references therein]{Fabian+12,King+15,Zubovas+17a,Trussler+20}. On the other hand, some models and simulations suggest that these outflows and jets can in some cases compress the galactic gas, and therefore enhance/trigger the star-formation \citep[e.g.][]{Rees+89,Hopkins+12,Nayakshin+12,Ishibashi+12,Zubovas+13,El-Badry+16,Bieri+16,Zubovas+13,Zubovas+17a,Wang+18,Gallagher+19,Bessiere+22}. Additionally, cosmological simulations \citep[e.g.][]{Springel+05,Vogelsberger+14,Crain+15} without the inclusion of feedback effects effects do not reproduce the galaxies luminosity function, underestimate the ages of the most massive galaxies \citep[see Figs.~8~and~10 of][]{Croton+06} and do provide a limited insight into the nature and source of the feedback processes \citep[e.g. AGN or SN dominated][]{Schaye+15}.

To better understand this kind of process it is very important to study the galaxies' properties in a spatially resolved way so that the AGN effects can be probed at different locations in the host galaxies. This can be done via the use of Integral Field Units (IFUs), which allow the study of ionization sources across the galaxies, the interplay among the global/integrated properties of galaxies with the local
(spatially resolved) ones, and the link between these two types of relations as well as the spatial (radial) distribution of such properties \citep{Sanchez+20,Sanchez+21}.

However, extracting information from IFU data can be complex \citep{Steiner+09}, and hence the analysis of such data with different techniques is desirable. Over the years, our AGNIFS team has developed a series of studies analysing IFU data \citep[e.g.][]{Storchi-Bergmann+09a,Storchi-Bergmann+12a,Mallmann+18,Schnorr-Muller+21,Riffel+11a,Riffel+17,Riffel+18,Riffel+21,Riffel+21b,Riffel+21a,Riffel+22,Ilha+22} and methods to analyze this type of data. 

In the present paper, we present {\sc megacubes} \citep{Mallmann+18} with our fitting procedures for the $\sim$ 10\,000 datacubes of the Mapping Nearby Galaxies at Apache Point Observatory (MaNGA) survey \citep{Bundy+15a}, as well as an analysis of the stellar content of the final AGN sample and a matched control sample \citep{Rembold+17}. This paper is structured as follows: in section~\ref{sec:data} we present the updated samples. The fitting procedures and the {\sc megacubes} are described in section~\ref{sec:fitting}. The results are presented and discussed in section~\ref{sec:res_disc}  and final remarks are made in section~\ref{sec:final_rem}. We have used throughout the paper $H_0=73$\,km\,s$^{-1}$\,Mpc$^{-1}$ \citep{Riess+22}.

\section{data }\label{sec:data}

The data used here are those provided by the fourth-generation Sloan Digital Sky Survey (SDSS IV) sub-project Mapping Nearby Galaxies at Apache Point Observatory \citep[MaNGA,][]{Bundy+15a}. The survey has provided optical IFU spectroscopy ($3600$\,{\AA}-$10400$\,{\AA}), high quality data, of $\sim 10,010$ nearby galaxies (with $\langle z\rangle\,\approx\,0.03$). The observations were carried out with fiber bundles of different sizes (19-127 fibers) covering a field of $12''$ to $32''$ in diameter. MaNGA observations are divided into ``primary'' and ``secondary'' targets, the former was observed up to $1.5$ effective radius ($R_e$) while the latter was observed up to $2.5\,R_e$. For more details, see \citet{Drory+15,Law+15,Yan+16a,Yan+16}. 

We have fitted the stellar population and emission lines for the 10\,010 high quality unique\footnote{We have discounted the Coma, IC342, M31, and globular cluster targets, and cubes with ``UNUSUAL" or ``CRITICAL" data quality flags, as well as repeated observations.} data cubes available in  the MaNGA final data release \citep[DR17, ][]{Abdurrouf+22,Aguado+19,Blanton+17,Bundy+15a,Belfiore+19,Westfall+19,Cherinka+19,Wake+17,Law+15,Law+16,Law+21,Yan+16,Yan+16a,Drory+15,Gunn+06,Smee+13}.

\subsection{Active Galactic Nuclei and Control Samples in MaNGA}\label{sec_sample}

In this work, we update and expand the sample of AGN hosts optically identified in the MaNGA survey by \citet{Rembold+17}.
We have cross-matched all galaxy data cubes observed in DR17 with the SDSS-III spectroscopic data from DR12 \citep{Alam+15}. The fluxes and equivalent widths of the emission lines 
H$\beta$, [O\,{\sc iii}]$\lambda$5007, H$\alpha$  and [N\,{\sc ii}]$\lambda$6583, measured in the
SDSS-III integrated nuclear spectrum, were drawn from \citet{Thomas+13}.
We then classified the galaxies according to the ionizing source of the gas
using both the [\ion{N}{ii}] based \citet[][BPT]{Baldwin+81} diagram\footnote{It is worth mentioning that this diagram identify LINERs that would be missed by other diagrams \citep{Agostino+21}}. and the WHAN diagram \citep{CidFernandes+10}, following \citet{Rembold+17}. A galaxy is confirmed as an optical AGN host if it is located simultaneously in the Seyfert/LINER region of both the BPT and the WHAN diagrams. We refer to \citet{Rembold+17} for more details on the selection of AGN hosts. This methodology results in 298 confirmed optical AGN hosts in MaNGA.

As in  \citet{Rembold+17}, we have defined a sample of non-active control galaxies comparable to the AGN hosts in terms of redshift, stellar mass and morphology. We selected from DR17, as potential control galaxies, those presenting or not detectable emission lines in their SDSS-III fiber spectra, except those already classified as AGN hosts (see above). Besides the objects lacking detectable emission lines altogether, potential control galaxies are those located, within the uncertainties, in the star-forming region of the BPT diagram, or in other regions of the BPT diagram if the WHAN diagram discards ionization from an AGN. These comprise therefore objects ionized by young stars or by hot, low-mass evolved stars (HOLMES). For each AGN host, we then select a preliminary subsample of galaxies from the list of potential control objects that match their redshifts and stellar masses at deviations lower than 30 percent deviation. Such preliminary galaxies (typically a couple hundred for each AGN host) were then visually inspected, and the two objects that best match the morphology and axial ratio of the AGN host were selected as their control ``partners''.
For five AGN hosts (MaNGA id 1-37440, 1-189584, 31-115, 1-641156 and 1-300461), no control ``partners'' were selected, either because their redshifts are too close to the redshift limits of the MaNGA survey (e.g. 1-189584 at $z=0.0037$) or because their morphologies are not reproduced by other objects in the survey (e.g. 1-641156).

Our final sample is therefore composed of 293 AGN hosts paired with two non-active control galaxies each.  The control sample comprises 492 unique objects, less than the expected 586, due to multiple occurrences of the same control galaxy as control ``partners'' of more than one AGN host. Table~\ref{tableagns} and Table~\ref{tablecontrol} list relevant parameters of the AGN hosts and control galaxies respectively.

\begin{figure*}
	\includegraphics[width=1.9\columnwidth]{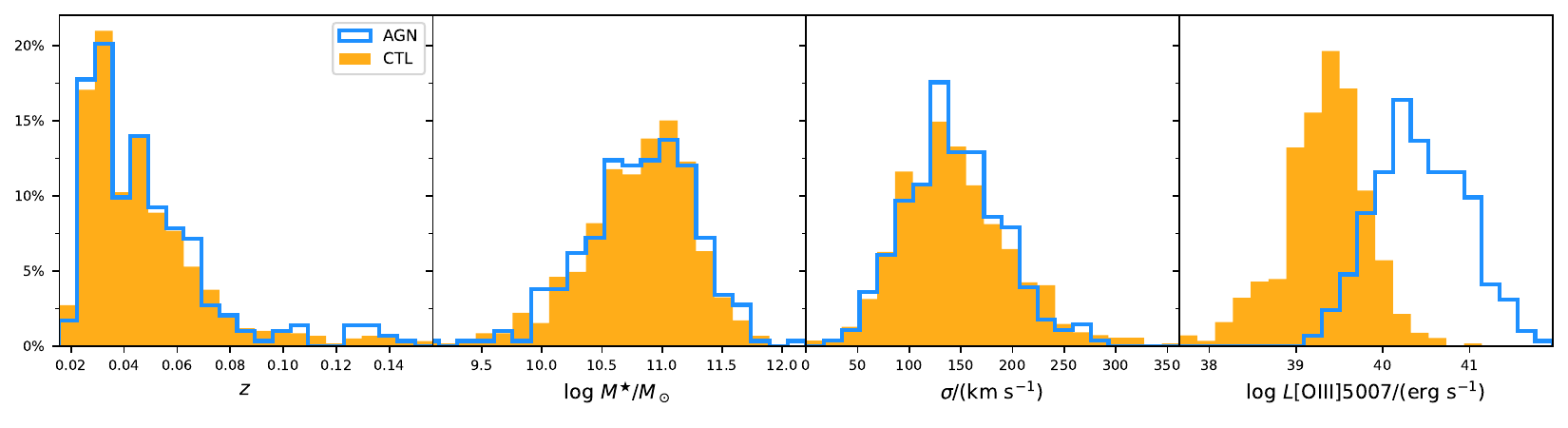}
    \caption{
    Distributions of redshift, stellar mass, stellar velocity dispersion, and [\ion{O}{iii}] luminosity of the AGN host sample (blue) and the control sample (orange). These properties are from SDSS single fiber observations. The AGN and the control sample present markedly different distributions of L([\ion{O}{iii}]), which is expected \citep[see][for a discussion]{Rembold+17}, but have other parameters similar between the two samples.          }
    \label{fig:sample}
\end{figure*}

As a verification of the quality of the control sample selection process, we confirmed using the Kolmogorov-Smirnov test that the distributions of stellar mass and redshift are indistinguishable between AGN hosts and control galaxies. We show the distribution of redshift, stellar mass, stellar velocity dispersion, and [\ion{O}{iii}] luminosity in Fig.~\ref{fig:sample}. Also, according to the Galaxy Zoo project \citep{Lintott+11}, 32\%, 60\% and 5\% of our AGN hosts are elliptical, spiral, and merging galaxies respectively. This morphological distribution is well matched by the control sample, comprised of 35\%, 61\% and 2\% of elliptical, spiral and merging galaxies respectively. We have also compared the numerical T-Type between AGN hosts and control galaxies from \citet{Vazquez-Mata+22} and found that both distributions are very similar (and indistinguishable as verified using the Kolmogorov-Smirnov test). For more details on the control sample selection see \citet{Rembold+17}.

\begin{table*}
\caption{Parameters of AGN in MaNGA-DR17. (1) galaxy identification in the MaNGA survey; (2)-(3): RA/DEC (2000) in degrees; (4) spectroscopic redshift from SDSS-III; (5): integrated
absolute $r$-band magnitude from SDSS-III; (6): logarithmic stellar mass in units of $M_\odot$; (7)-(8) morphological classification in the Hubble scheme and numerical T-Type from \citet{Vazquez-Mata+22}; (9): [OIII] luminosity in units of $10^{40}$\,erg\,s$^{-1}$. We only present here the first 6 lines of this table; the full version is available with the electronic version of the paper.}
\begin{tabular}{ccccccccc}
\hline
mangaID & RA  & DEC & $z$ & $M_r$ & log $M^\star/M_\odot$ & Type & T-Type & $L(\rm{[OIII]})$\\
 (1)    & (2)      & (3) & (4) & (5) & (6)   & (7)                   & (8)   & (9)  \\
\hline
1-558912 & 166.129410 & 42.624554 & 0.1261 & -20.46 & 11.25 & E & -5 & 56.82$\pm$1.25\\
1-150947 & 183.263992 & 51.648598 & 0.0849 & -20.23 & 10.96 & Sb & \phantom{-}3 & 53.82$\pm$1.75\\
1-39376 & \phantom{1}11.999101 & 13.742678 & 0.0567 & -19.76 & 10.92 & SBb & \phantom{-}3 & 43.88$\pm$0.65\\
1-295542 & 246.255981 & 24.263155 & 0.0503 & -19.34 & 10.73 & Edc & -5 & 39.10$\pm$0.87\\
1-458092 & 203.190094 & 26.580376 & 0.0470 & -18.08 & \phantom{1}9.64 & E & -5 & 35.30$\pm$1.04\\
1-211165 & 248.628647 & 37.695442 & 0.0991 & -21.04 & 11.24 & S0 & -2 & 33.83$\pm$0.57\\
\hline
\end{tabular}
\label{tableagns}
\end{table*}

\begin{table*}
\caption{Control sample parameters. (1) identification of the AGN host associated to the control galaxy; (2)-(10) same as (1)-(9) of Table~\ref{tableagns} for the control sample. We only present here the first 12 lines of this table; the full version is available with the electronic version of the paper.}
\begin{tabular}{cccccccccccccc}
\hline
AGN mangaID & mangaID & RA  & DEC & $z$ & $M_r$ & log $M^\star/M_\odot$ & Type & T-Type &  $L(\rm{[OIII]})$\\
 (1)    & (2)      & (3) & (4) & (5) & (6)   & (7)                   & (8)     & (9)  & (10) \\
\hline
1-558912 & 1-71481 & 117.456001 & 34.883911 & 0.1312 & -20.95 & 11.70 & E & -5 & \phantom{1}0.10$\pm$0.20\\
 & 1-72928 & 127.256485 & 45.016773 & 0.1270 & -20.62 & 11.52 & E & -5 & \phantom{1}0.09$\pm$0.23\\
1-150947 & 1-338555 & 114.374146 & 41.737232 & 0.0893 & -17.42 & \phantom{1}8.99 & E & -5 & \phantom{1}2.05$\pm$0.39\\
 & 1-248214 & 240.347031 & 43.085041 & 0.0600 & -19.80 & 10.97 & Edc & -5 & \phantom{1}0.41$\pm$0.14\smallskip\\
1-39376 & 1-285052 & 199.061493 & 47.599365 & 0.0573 & -19.77 & 10.85 & SBab & \phantom{-}2 & \phantom{1}0.11$\pm$0.03\\
 & 1-283662 & 191.217728 & 42.422195 & 0.0536 & -19.27 & 10.66 & SB0a & \phantom{-}0 & \phantom{1}0.00$\pm$0.00\\
1-295542 & 1-94228 & 248.018265 & 47.764759 & 0.0494 & -19.19 & 10.58 & Sa & \phantom{-}1 & \phantom{1}0.30$\pm$0.06\\
 & 1-92774 & 243.407715 & 49.069408 & 0.0552 & -19.28 & 10.75 & Edc & -5 & \phantom{1}0.39$\pm$0.23\smallskip\\
1-458092 & 1-38319 & \phantom{1}51.609463 & -0.311706 & 0.0376 & -17.24 & \phantom{1}9.76 & Edc & -5 & \phantom{1}0.00$\pm$0.00\\
 & 1-323888 & 245.686630 & 32.659161 & 0.0410 & -18.64 & \phantom{1}9.54 & S0 & -2 & 11.66$\pm$0.31\\
1-211165 & 1-55272 & 146.661270 & \phantom{1}2.658917 & 0.0933 & -20.90 & 11.49 & S0 & -2 & \phantom{1}1.24$\pm$0.27\\
 & 1-262966 & 229.138367 & 32.243935 & 0.0913 & -20.79 & 11.22 & Edc & -5 & \phantom{1}0.16$\pm$0.16\smallskip\\
\hline
\end{tabular}
\label{tablecontrol}
\end{table*}

\section{Fitting procedures}\label{sec:fitting}

\subsection{Stellar Population}
In this section, we describe the procedures adopted in the stellar population fits.

\subsubsection{Starlight} \label{sec:starlight}

To perform a full spectral fitting stellar population synthesis on our datacubes we employed the \st\ code \citep{CidFernandes+05,CidFernandes+18} which combines the spectra of a {\it base of elements} of $N_{\star}$ simple stellar population (SSP) template spectra $b_{j,\lambda}$, weighted in different proportions, in order to reproduce the observed spectrum $O_\lambda$. For this modeling, the observed and modeled spectra $M_\lambda$ are normalized at a user-defined wavelength $\lambda_0$. The reddening is given by the term $r_\lambda = 10^{-0.4 (A_\lambda - A_{\lambda_0})}$, weighted by the population vector $x_j$ (which represents the fractional contribution of the $j$th SSP to the light at the normalisation wavelength $\lambda_0$), and convolved with a Gaussian distribution $G(v_\star, \sigma_\star)$ to account for velocity shifts $v_{\star}$, and velocity dispersion $\sigma_{\star}$. Each model spectra can be expressed as:

\begin{equation}
	M_\lambda = M_{\lambda_0} \left[ \sum_{n=1}^{N_\star} x_j\,b_{j,\lambda}\,r_\lambda \right] \otimes G(v_\star, \sigma_\star),
\end{equation}

\noindent where $M_{\lambda_0}$ is the flux of the synthetic spectrum at the wavelength $\lambda_0$.  To find the best parameters for the fit, the code searches for the minimum of $\chi^2 = \sum_{\lambda_i}^{\lambda_f} [(O_\lambda - M_\lambda) \omega_\lambda]^2$, where $\omega_\lambda$ is the inverse of the error, using a simulated annealing plus Metropolis scheme. 

The {\it base of elements}  we use is the $GM$ described in \citet{CidFernandes+13,CidFernandes+14} that is constructed using the {\sc Miles} \citep{Vazdekis+10} and \citet{GonzalezDelgado+05} models. We have updated it with the {\sc Miles} V11 models \citep{Vazdekis+16a}.  We used 21 ages (t= 0.001, 0.006, 0.010, 0.014, 0.020, 0.032, 0.056, 0.1, 0.2, 0.316, 0.398, 0.501 0.631, 0.708, 0.794, 0.891, 1.0, 2.0, 5.01, 8.91 and 12.6 Gyr) and four metallicities (Z= 0.19, 0.40, 1.00 and 1.66 Z$_\odot$). Following \citet{CidFernandes+04} we have also added to the {\it base of elements} a power law of the form $F_{\nu} \propto \nu^ {-1.5}$ to account for the contribution of a possible AGN featureless continuum (FC), observed directly or as scattered light \citep[see also][for discussion on the FC contribution]{Koski+78,CidFernandes+05,Martins+13,Riffel+09,Riffel+21,Riffel+22}. It is worth mentioning that FC is  only considered in the light fractions being removed from the calculations of stellar properties (e.g. mean ages and metallicities, see \S~\ref{megacubemod}). 

Finally, the normalization flux at $\lambda_0$ was adopted to be the mean value between 5650\AA\ and 5750\AA. The reddening law we used was that of \citet{Cardelli+89} and the synthesis was performed in the spectral range from $3700$\,{\AA} to $6900$\,{\AA}.

\subsubsection{Urutau}

\st\ is not developed to work with datacubes or {\it fits } files, therefore we have developed an in house software: {\sc urutau}. It was developed in Python to handle input and output data from multiple sources via a modular pipeline execution. The modules of the pipeline can be swapped depending on the user case (such as datacube extraction method, spectral fitting software, desired scientific result, etc.). {\sc Urutau} was also planned with parallel processing capability in order to execute many instances of a closed source executable (such as {\sc starlight}) for different sets of data. The code is available at: \url{https://github.com/ndmallmann/urutau}.

\subsubsection{Stellar population {\sc megacube} modules}\label{megacubemod}

To improve the management of the {\sc starlight} inputs and outputs, the organization of the results and the analysis, we developed a series of modules to run under {\sc urutau} and appended the data-products to the original datacube. These modules are as follows:

{a) \sc libMaNGA}: This library is used to prepare and convert the MaNGA datacubes to the format expected by \st. The main steps are as follows \citep{Mallmann+18} :
\begin{enumerate}
\item Filtering of the datacubes using a two-dimensional butterworth filter to remove spurious data and increase the signal-to-noise ratio. It does not require any addition of adjacent spaxel, thus allowing to a better exploration of the spatial resolution. A better description of this technique  can be found in \cite{Riffel+16};

\item Correct each spaxel for Galactic reddening  using the \citet{Schlegel+98} extinction maps and the \citet{Cardelli+89} reddening law;

\item Correct for redshift using the SDSS-III redshift values available in the {\it drpall} tables of the MaNGA database;

\item Estimation of the signal-to-noise ratio (SNR) maps in the wavelength range 5650-5750\,{\AA} for every spaxel. We provide masks with cuts of SNR values of 3, 5, and 10, as well as the SNR map. We have performed our fits for spaxels with SNR$\ge$1. Note, however, that any cutoff in SNR larger than 1 can be applied by the user since we provide the SNR maps.

\end{enumerate}

{b) \sc libStarlightCaller:} This module is used to actually run \st\ with the configuration listed in \S~\ref{sec:starlight}. In other words, it sets up all the configuration files needed for the fits and runs \st\ for each individual spaxel.

{c) \sc libCreateMegacube:} With this function we derive additional parameters like: mean ages and metallicities, star formation rates, and binned population vectors from the starlight output as follows:

\begin{description}
    \item[\it Binned population vectors:] Since small differences in ages are washed away due to noise effects, one coarser, but more robust way is to define binned population vectors \citep[e.g.][]{CidFernandes+05,Riffel+09}. We included in our {\sc megacubes} the following binned population vectors (where y=young, i=intermediate, and o=old): 
    \begin{description}
     \item[FC1.50:] The percent contribution (at $\lambda_0$) of a featureless component of the form $F_{\nu} \sim \nu^{-1.5}$; 
     
     \item[xyy\_light (mass):] Light (mass) binned population vector in the age range  $t \leq$ 10~Myr;
     
     \item[xyo\_light (mass):] Light (mass) binned population vector in the age range 14~Myr $ <  t \leq$ 56~Myr;
     
     \item[xiy\_light (mass):] Light (mass) binned population vector in the age range 100~Myr $ <  t \leq$ 500~Myr;
     
     \item[xii\_light (mass):] Light (mass) binned population vector in the age range 630~Myr $ <  t \leq$ 800~Myr;
     
     \item[xio\_light (mass):] Light (mass) binned population vector in the age range 890~Myr $ <  t \leq$ 2.0~Gyr;
     
     \item[xo\_light (mass):]  Light (mass) binned population vector in the age range 5.0~Gyr $ <  t \leq$ 13~Gyr;

    \end{description}

    \item[\it Star Formation Rates via stellar population fit:] the star formation rate obtained from the stellar population fit ($SFR_\star$) over an user-defined age interval ($\Delta t = t_{j_f}-t_{j_i}$) is also computed in this module. This can be computed since the SSPs model spectra are in units of L$_{\odot}$ \AA$^{-1} M_{\odot}^{-1}$, and the observed spectra ($O_{\lambda}$) are in units of $\rm erg/s/cm^2/$\AA. The SFR$_\star$ over the chosen $\Delta t$  can be computed assuming that the mass of each base component ($j$) which has been processed into stars can be obtained as:
\begin{equation}
    M_{\star,j}^{\rm ini} = \mu_{j}^{\rm ini} \times \frac{4\pi d^2}{3.826\times 10^{33}},
\end{equation}
where $M_{\star,j}^{\rm ini}$ is given in $M_\odot$, $\mu_{j}^{\rm ini}$ is associated with the mass that has been converted into stars for the $j$-th element and its flux. This parameter is given in $M_\odot\,{\rm erg s^{-1} cm^{-2}}$; $d$ is the luminosity distance to the galaxy in cm and 3.826$\times10^{33}$ is the Sun's luminosity in erg\,s$^{-1}$. Thus, the SFR over the $\Delta t$ as defined above can be obtained from the equation:
\begin{equation}
    {\rm SFR_\star} = \frac{\sum_{j_i}^{j_f} M_{\star,j}^{\rm ini}}{\Delta t}.
\end{equation}
 For more details see \citet{Riffel+21}. Here we computed the $ {\rm SFR_\star}$ over the last ($\Delta t$) 1\,Myr, 5.6\,Myr, 10\,Myr, 14\,Myr, 20\,Myr, 32\,Myr, 56\,Myr, 100\,Myr, and 200\,Myr (they were labelled as SFR\_1, SFR\_5, SFR\_10, SFR\_14, SFR\_20, SFR\_30, SFR\_56, SFR\_100, and SFR\_200, respectively). 

\item[\it Mean ages and metallicities:] Following \citet[][]{CidFernandes+05} we have computed the mean ages (the logarithm of the age, actually) for each spaxel weighted by the stellar light 
\begin{equation}\langle {\rm log} t_{\star} \rangle_{L} = \displaystyle \sum^{N_{\star}}_{j=1} x_j {\rm log}t_j, \end{equation}
and weighted by the stellar mass, 
\begin{equation}\langle {\rm log} t_{\star} \rangle_{M} = \displaystyle \sum^{N_{\star}}_{j=1} \mu_j {\rm log}t_j. \end{equation}
The light-weighted mean metallicity is defined as  
\begin{equation}\langle Z_{\star} \rangle_{L} = \displaystyle \sum^{N_{\star}}_{j=1} x_j Z_j,\end{equation}
and the mass-weighted mean metallicity is defined by:
\begin{equation}\langle Z_{\star} \rangle_{M} = \displaystyle \sum^{N_{\star}}_{j=1} \mu_j Z_j.\end{equation} 

Note that both definitions are limited by the age and metallicities range used in our elements base. They are presented under the  {\it Mage\_L, Mage\_M, MZ\_L,} and {\it MZ\_M} keywords. 

\item[\it Additional parameters:] we also present a set of additional fitting parameters that are included in  the {\sc megacubes}, such as 
optical extinction (Av), 
the present mass in stars (M*), 
masses processed in stars (M*in),
normalisation flux (F\_Norm),
stellar dispersion velocity (Sigma\_star),
stellar rotation velocity (vrot\_star),
percentage mean deviation (Adev),
reduced $\chi^2$ (ChiSqrt), and
signal-to-noise ratio on normalisation window (SNR).  
    
\end{description}

The final product after the procedures described above is a complete set of Flexible Image Transport System ({\sc fits}) files for the entire MaNGA sample; this constitutes what we refer to as {\sc megacubes}.
The results obtained after our fitting are appended as multiple extensions into the original MaNGA datacubes. The additional extensions added are as follows: 
\begin{description}
 \item[\bf BaseAgeMetal:] The ages and metallicities of the SSPs used in the base of elements.
 
 \item[\bf PopBins:] In this extension we store the synthesis parameters directly computed  by \st\ as well as properties derived in the {\sc libCreateMegacube} module (see \S~\ref{megacubemod})

\item[\bf PoPVecsL (M):] Original (e.g. not binned) population vectors in light fractions (mass fractions).
\end{description}

The \str\ output observed flux (FLXOBS), synthetic flux (FLXSYN), and weights (WEIGHT), as well as continuum signal-to-noise ratio masks, are also stored on the {\sc megacubes}.

\subsection{Emission line fitting}

Our {\sc megacube}s, by construction, include absorption-free emission-line datacubes. We have used these datacubes to fit the most common emission lines in the optical region. To this purpose we use the {\sc ifscube}\footnote{https://github.com/danielrd6/ifscube} Python package \citep{Ruschel-Dutra+20,Ruschel-Dutra+21} to fit the profiles of the most prominent emission-line, namely: H$\beta$, [O\,{\sc iii}]$\lambda\lambda$4959,5007, He\,{\sc i}$\lambda$5876, [O\,{\sc i}]$\lambda$6300, H$\alpha$ [N\,{\sc ii}]$\lambda\lambda$6548,6583 and [S\,{\sc ii}]$\lambda\lambda$6716,6731.  The line profiles are fitted with Gaussian curves by adopting the following constraints: (i) the width and centroid velocities of emission lines from the same parent ion are constrained to the same value, just He\,{\sc i}$\lambda$5876 and [O\,{\sc i}]$\lambda$6300 are kept free during the fit; (ii) the [O\,{\sc iii}]$\lambda$5007/$\lambda$4959 and  [N\,{\sc ii}]$\lambda$6583/$\lambda$6548 flux ratios are fixed to their theoretical values of 2.98 and 3.06, respectively; (iii) the centroid velocity is allowed to vary from --300 to 300 km\,s$^{-1}$ for [\ion{S}{ii}] lines and --350 to 350 km\,s$^{-1}$ for the other lines relative to the velocity obtained from the redshift of each galaxy (listed in the MaNGA Data Analysis Pipeline - DAP); and (iv) the observed velocity dispersion of all lines is limited to the range 40--300 km\,s$^{-1}$. In a few sources better fits of the observed profiles are obtained when including a broad component for \ha\ and \hb\footnote{No changes in the activity classification are found when a broad component is included since, even in cases where a broad component is evident, the line fluxes are well represented by a single component. For completeness, we have made available the {\sc megacubes}  for these sources with and without the broad component.
}. In addition, we include a first-order polynomial to reproduce the local continuum.

Instead of relying on the minimization of a residual function, kinematic constraints were enforced by construction, as each emission line in the same kinematic group inherits the same exact parameter for line width and centroid velocity, see \citet{Ruschel-Dutra+21} for additional details. Additionally, all the lines have been fitted with a single Gaussian. This approach has the advantage of reducing the number of free parameters that are perceived by the fitting algorithm.

The output of IFSCube is in the form of a Multi-Extension FITS file (MEF), with each extension storing a different result from the fitting process, a copy of the input, or software-specific parameters for later reference.
The first four extensions are in the same shape of the input data, "FITSPEC" representing the observed spectra (restricted to the requested wavelength window), the result of the pseudo continuum fit is in "FITCONT", "STELLAR" stores the given stellar spectra and "MODEL" stores the modelled emission spectrum.
If no stellar spectra were given as input, the corresponding extension is filled with zeros.

Next there is the "SOLUTION" extension which stores the best parameters for all the emission lines considered, and also the reduced chi-squared of the model as a whole.
The shape is that of the input data for the spatial part, with each parameter occupying a different plane.
For instance, if three Gaussian curves were fit to a 50x50 data cube, then there should be 9 independent parameters for the whole model, and another plane for the chi-squared, resulting in a 50x50x10 array.

The initial guess for each parameter is stored in the "INIGUESS" extension, also in the form of a three-dimensional array, with a set of parameters for each spaxel.
This is useful only when using the option to update the initial guess based on results from previous spaxels.
Extensions 7 through 10 ("FLUX\_D", "FLUX\_M", "EQW\_M" and "EQW\_D") store the fluxes and equivalent widths of each line, using two different approaches.
The first one is a direct trapezoidal integration of the observed spectra, minus the stellar component and pseudo-continuum, while the second is an integration over the modelled emission line.
The final status of each spaxel is specified by an integer code and stored in the "STATUS" extension.
A bi-dimensional mask, which combines both the input mask and IFSCube's internal assessment of the spaxel quality, is stored in the extension "MASK2D".

The last three extensions are tables, as opposed to all the previous ones which consist of images.
Extension "SPECIDX" stores the sequence in which the spaxels were processed.
Parameter names in the exact order that they apper in the "SOLUTION" extension are specified by the table in the "PARNAMES" extension.
Finally, the "FITCONFIG" extensions keeps a copy of the input configuration.

All these extensions regarding the emission line fitting have been appended to the {\sc megacubes}.

\section{Results and discussion} \label{sec:res_disc}

\subsection{{\sc megacubes}}

As described in the previous sections we have fitted the stellar population and derived their properties for $\sim$ 10\,000 galaxies, being thus the main purpose of this paper to make available for the community, in an easy-to-use way, the maps of the stellar population and emission lines of all spaxels on the $\sim$10 thousand MaNGA datacubes. This final product is what we call {\sc megacubes}.

In order to facilitate the inspection of the large number of properties that can be derived from the analysis of the spectra available in the  IFU data cube, a customized viewer was developed in collaboration with LIneA’s IT team. All the {\sc megacubes} are available through a web interface (\url{https://manga.linea.org.br/} or \url{http://www.if.ufrgs.br/~riffel/software/megacubes/}), where each one of them can be downloaded and/or inspected via a number of different plots, which can be interacted with. Additionally, we do also provide a table with the mean or integrated values (integrated for $M\star$ and emission-line fluxes) of the circularised maps for many properties for all the galaxies for different radial values as follows\footnote{We call attention to the equivalent width values since they are the mean value of the different strengths measured over the field. They should be taken as an upper limit: 
spaxels containing emission lines below the detectability limit contribute only to the continuum, thus reducing the total line flux, while in an integrated spectrum over the same field these ``spaxels'' contribute also
to the line flux.}:
{\sc r\_int\_0} = R$< 0.5\,R_e$;  
{\sc r\_int\_1} = $0.5\,R_e \leq$ R$< 1.0\,R_e$; 
{\sc r\_int\_2} = $1.0\,R_e \leq$ R$< 1.5\,R_e$, 
{\sc r\_int\_3} = $1.5\,R_e \leq$ R$< 2.0\,R_e$;
{\sc r\_int\_4} = R$< 0.5\,{\rm kpc}$;  
{\sc r\_int\_5} = $0.5\, {\rm kpc} \leq$ R$< 1.0\,{\rm kpc}$;
{\sc r\_int\_6} = $1.0\, {\rm kpc} \leq$ R$< 2.0\,{\rm kpc}$;
{\sc r\_int\_7} = $2.0\, {\rm kpc} \leq$ R$< 5.0\,{\rm kpc}$, and
{\sc r\_int\_8} = $5.0\, {\rm kpc} \leq$ R$< 10.0\,{\rm kpc}$. These mean/integrated properties have been derived considering the values where the continuum SNR$\ge$10 ({\sc SN\_MASKS\_10}) and,  for the emission-line related properties we have considered an additional threshold of $\ge 3\sigma$ detection.

As a highlight of the science that can be done with these {\sc megacubes}, we analyze here the stellar populations and the spatially resolved emission-line diagnostic diagrams of the final set of AGNs on the MaNGA survey and compare them with the results obtained for the control sample we have described in \S~\ref{sec_sample}.

\subsection{Mapping the stellar populations in MaNGA AGNs}

\begin{figure*}
	\includegraphics[width=1.9\columnwidth]{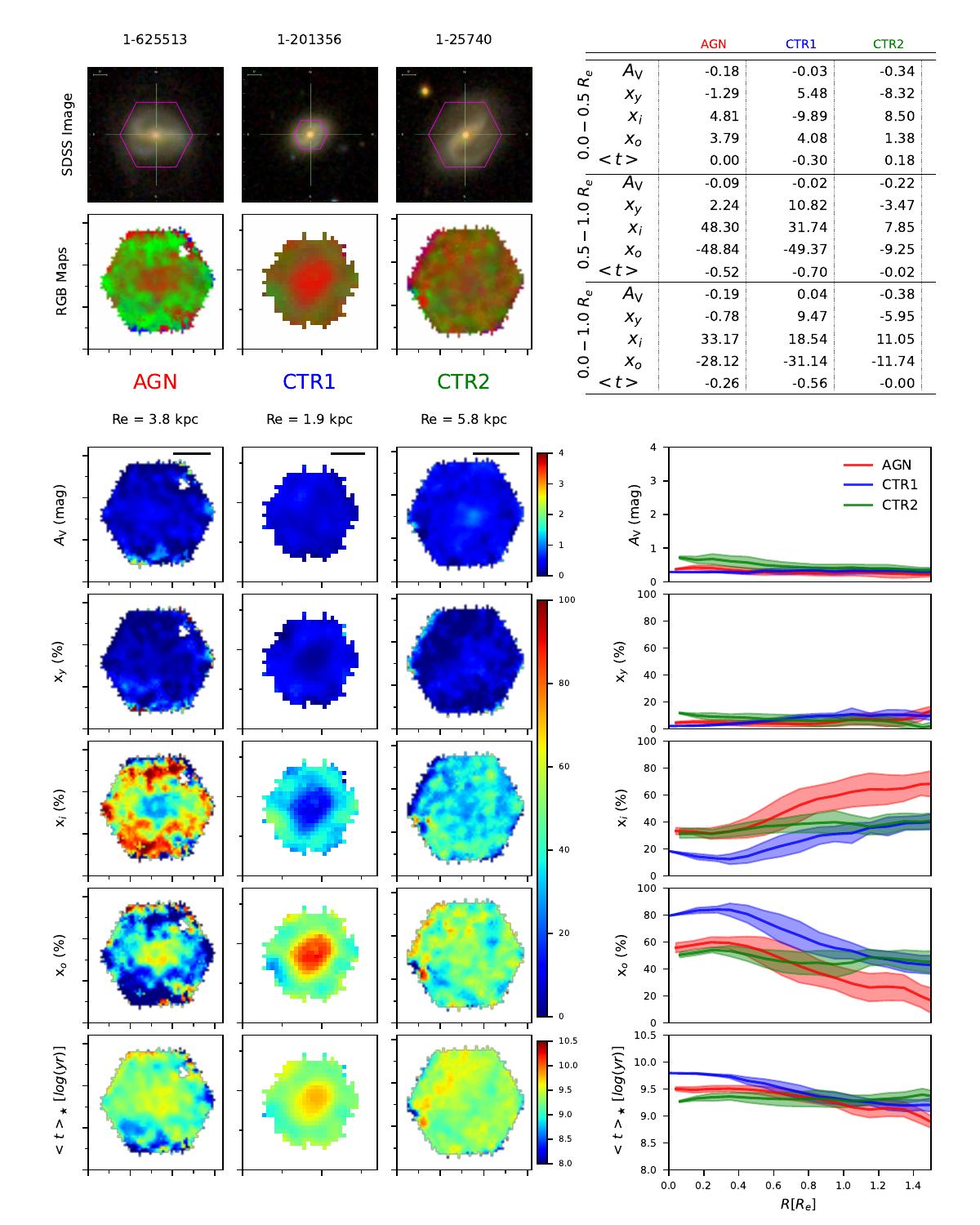}
    \caption{
    	Comparison of the stellar population properties between a late-type AGN and its controls. {\bf Left side panels -} {\it Top set of panels:} SDSS image (the MaNGA field is indicated in magenta). Second row:  RGB image using a composition of the binned population vectors [blue: young (xyy+xyo = xy); green: intermediate age (xiy+xii+xio = xi); red: old (xo)]. {\it Bottom set of panels:} From top to bottom: visual extinction ($A_{\rm V}$), $\rm X_Y$, $\rm X_I$, $\rm X_O$ and mean age ($<t>$) maps. For display purposes, we used tick marks separated by  5$"$. The solid horizontal line in the $A_{\rm V}$ maps represents 1\,$R_e$.
	{\bf Right side panels -} {\it Top:} summary table with the mean gradient values for each property in 3 different $R_e$ ranges. {\it Bottom:} average radial profiles for AGN hosts (red) and control galaxies (blue and green). The units of these profiles are the same as in the maps on the left side. The shaded area represents $1\,\sigma$ standard deviation. This figure is similar to those of \citet{Mallmann+18}, but the profiles presented here have been calculated using  new circularised deprojected maps. 
	      }
    \label{late-comp}
\end{figure*}

\begin{figure*}
	\includegraphics[width=1.9\columnwidth]{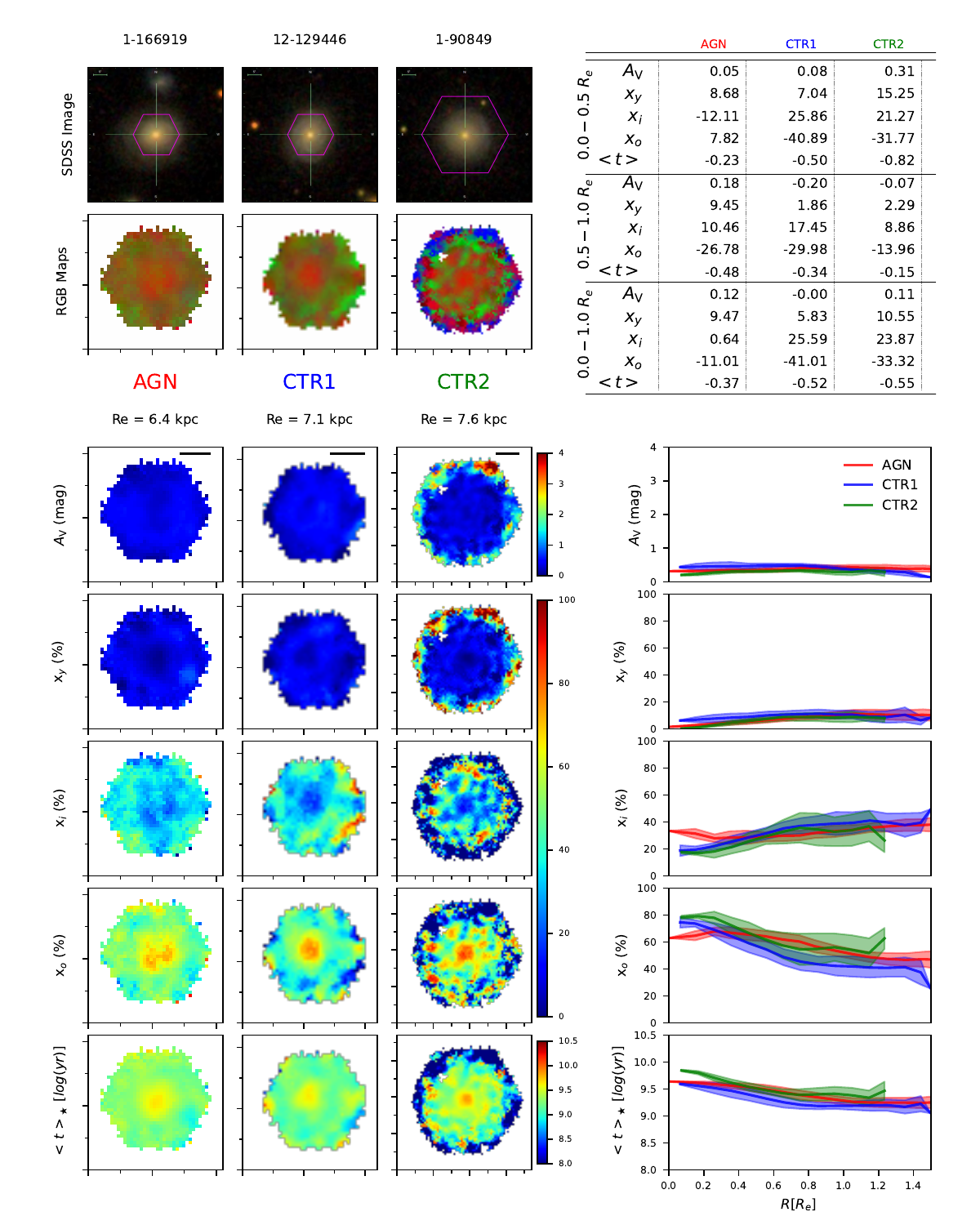}
    \caption{
    	Same as Fig.~\ref{late-comp} but for an early-type AGN and its control galaxies. 
     }
    \label{early-comp}
\end{figure*}

The {\sc megacubes} have a large number of output parameters, but here we are mostly interested in $x_{j}$ -- the fractional contribution of each SSP to the total light at  $\lambda_{norm}$ -- as they represent the star formation history of the galaxy. Following \citet{Mallmann+18}, we show example maps of these parameters, together with the mean age and the reddening maps in Fig.~\ref{late-comp} (late-type AGN) and Fig.~\ref{early-comp} (early-type AGN). It is worth mentioning that we have changed the methodology applied in \citet{Mallmann+18} for the computation of the profiles using a circularized deprojection over the entire FoV, instead of only a region along the major axis of the galaxies as in \citet{Mallmann+18}.

\subsubsection{2D maps analysis}
In order to show in a qualitative way how the different stellar population components are distributed along the FoV we have constructed RGB images (Figs.~\ref{late-comp}~and~\ref{early-comp}) of the galaxies that were created by assigning three colors (Red, Green, and Blue) to the binned population vectors: Red represents the old (xo: 5~Gyr $ <  t \leq$ 13~Gyr), green the intermediate age (xiy+xii+xio = xi: 100~Myr $ <  t \leq$ 2~Gyr), and blue the young (xyy+xyo = xy: $t \leq$\,56Myr) stellar populations (see \S~\ref{megacubemod} for the age bin definitions). We also show (left side) the maps for the mean properties and the radial profiles (right side) of these properties. In addition, we show the mean gradient values in 3 different radial steps (0.0 to 0.5R$_e$, 0.5 to 1.0R$_e$, and 0.0 to 1.0 R$_e$). These gradients have been computed by linear regression over the points within the radial steps and the values quoted are the angular coefficients for the AGN and for the two controls galaxies. 

From this exercise, one can see that the old population is more centrally concentrated in the late-type sources, while the gradient is shallower in the early-type ones, but in both cases, a decreasing value is observed with increasing radius. In addition, both types show significant contributions of the intermediate age  with an outwards increasing gradient, while the young component is distributed along the full FoV. This behavior is seen for almost all sources in our sample (see supplementary material for the figures for the other galaxies).  

\subsubsection{Gradients}

\begin{figure*}
	\includegraphics[width=2.\columnwidth]{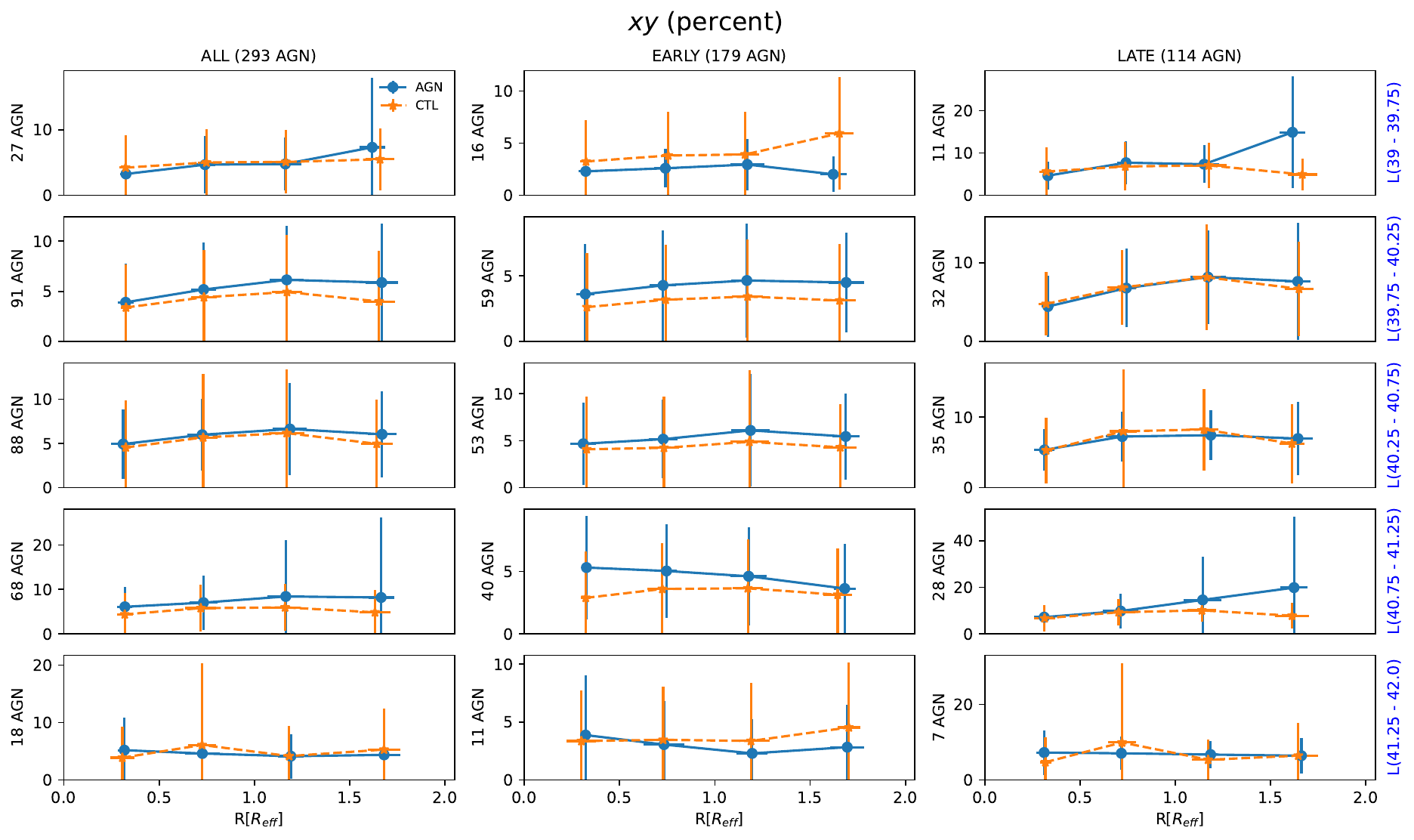}
    \caption{Comparison of the percent contribution to the light at $\lambda_0$=5700\AA\ between AGN (blue) and controls (orange) of the young stellar population (xyy+xyo; $t \leq$ 56~Myr) percent contribution (in light fractions) for five different luminosity bins (right side: $39$-$39.75$, $39.75$-$40.25$, $40.25$-$40.75$, $40.75$-$41.25$, $41.25$-$42.0$) as a function of the radius expressed in units of $R_e$. The second and third columns show also the grouping of the galaxies in early and late-type hosts.
    }
    \label{xyy}
\end{figure*}

\begin{figure*}
 	\includegraphics[width=2.\columnwidth]{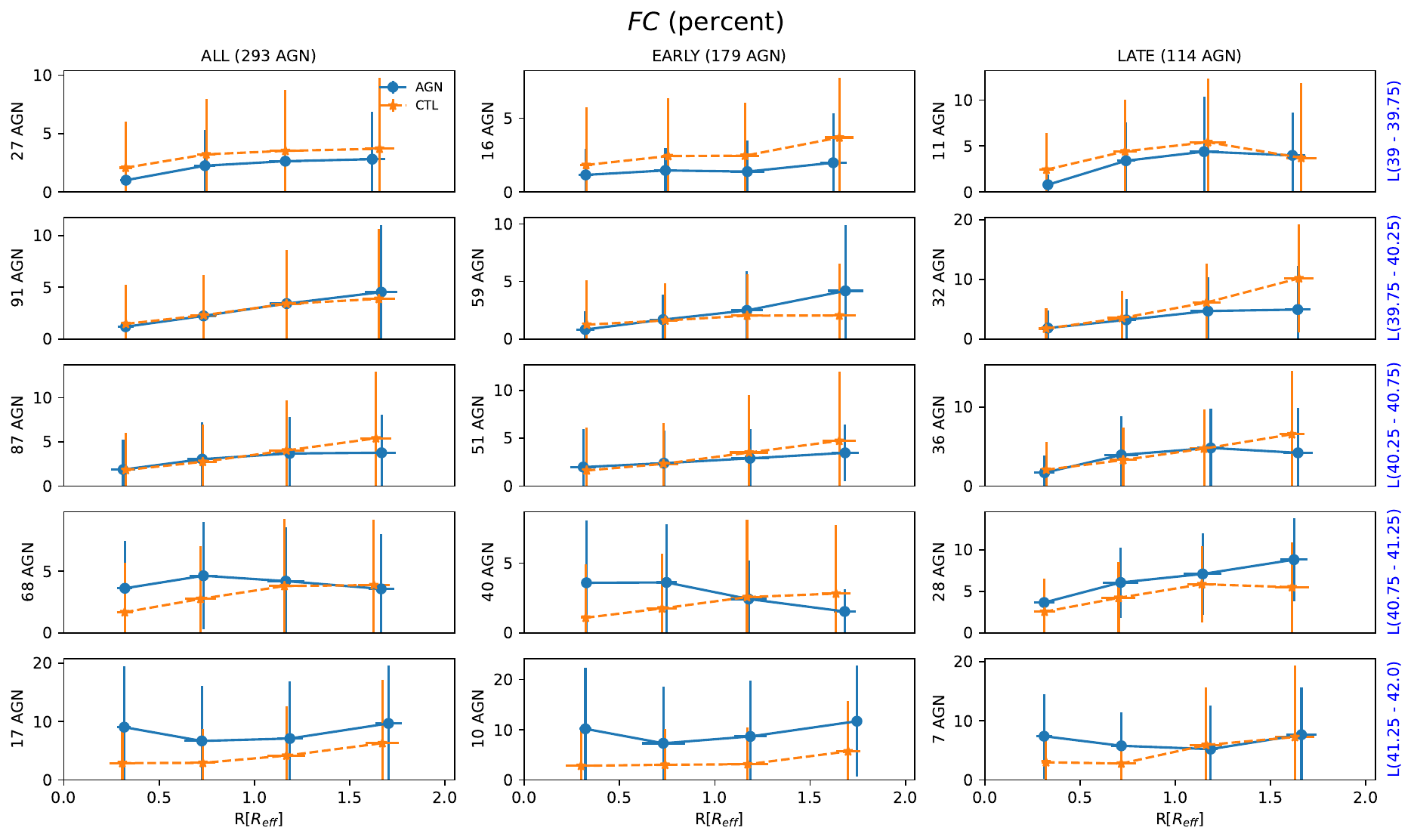}
\caption{The same as in Fig.~\ref{xyy} but for the featureless continuum component.} 
 \label{fc}
 \end{figure*}

\begin{figure*}
 \includegraphics[width=2.\columnwidth]{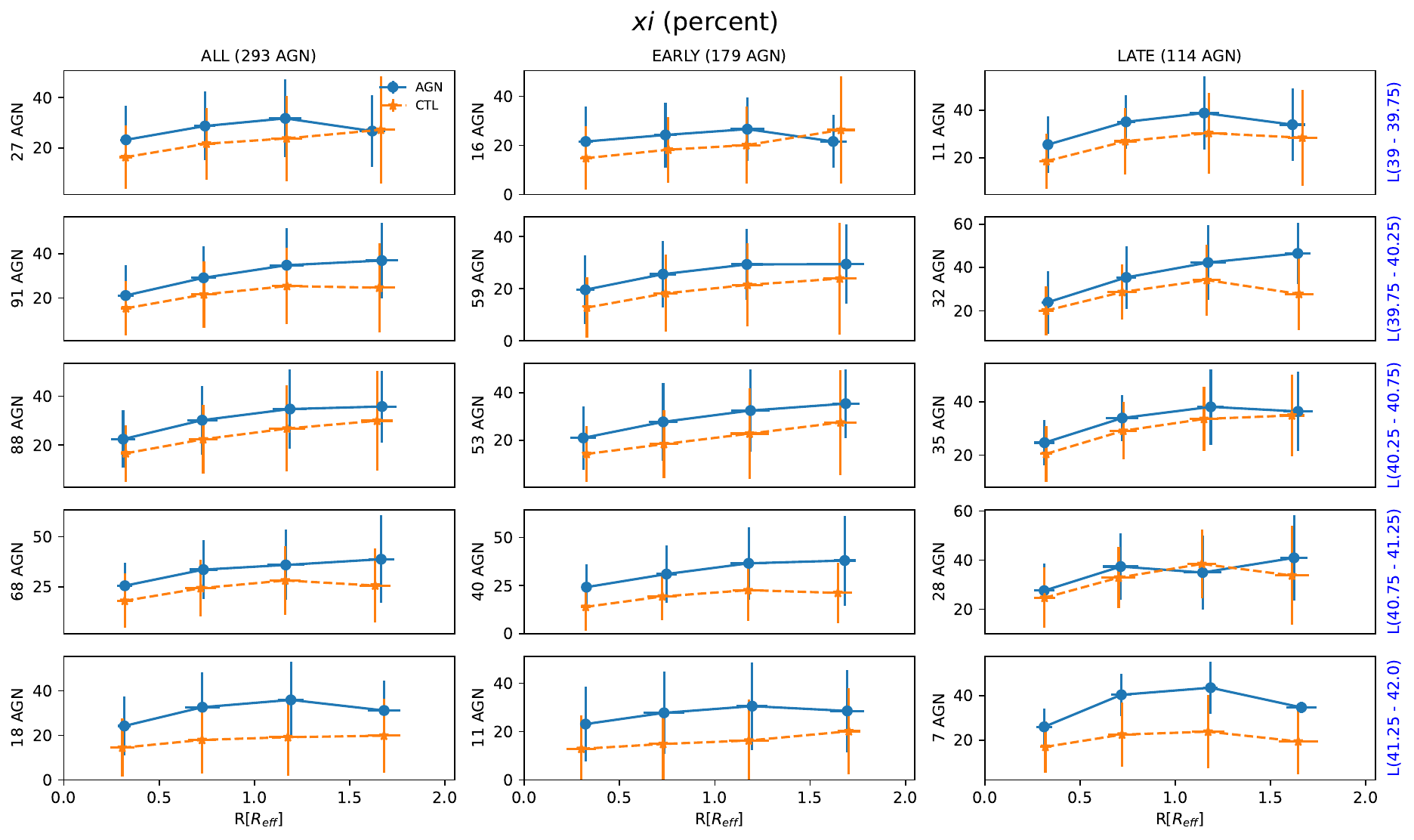}  
\caption{The same as in Fig.~\ref{xyy} but for the intermediate age stellar population (xiy+xii+xio; 100~Myr $ <  t \leq$ 2~Gyr) component.} 
 \label{xi}
 \end{figure*}

\begin{figure*}
 	\includegraphics[width=2.\columnwidth]{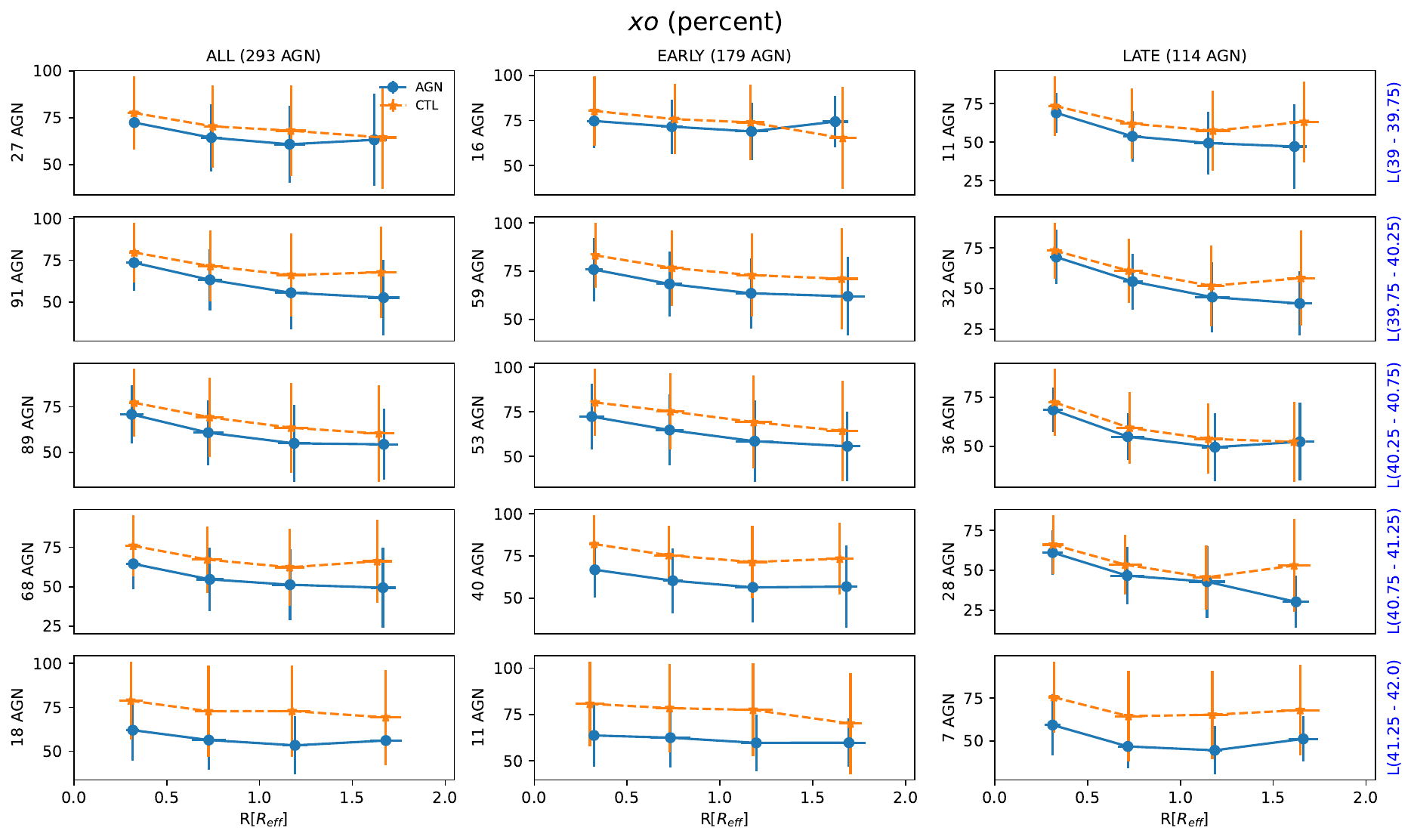}
\caption{The same as in Fig.~\ref{xyy} but for the xo (5~Gyr $ <  t \leq$ 13~Gyr) stellar population component.} 
 \label{xo}
 \end{figure*}

\begin{figure*}
 	\includegraphics[width=2.\columnwidth]{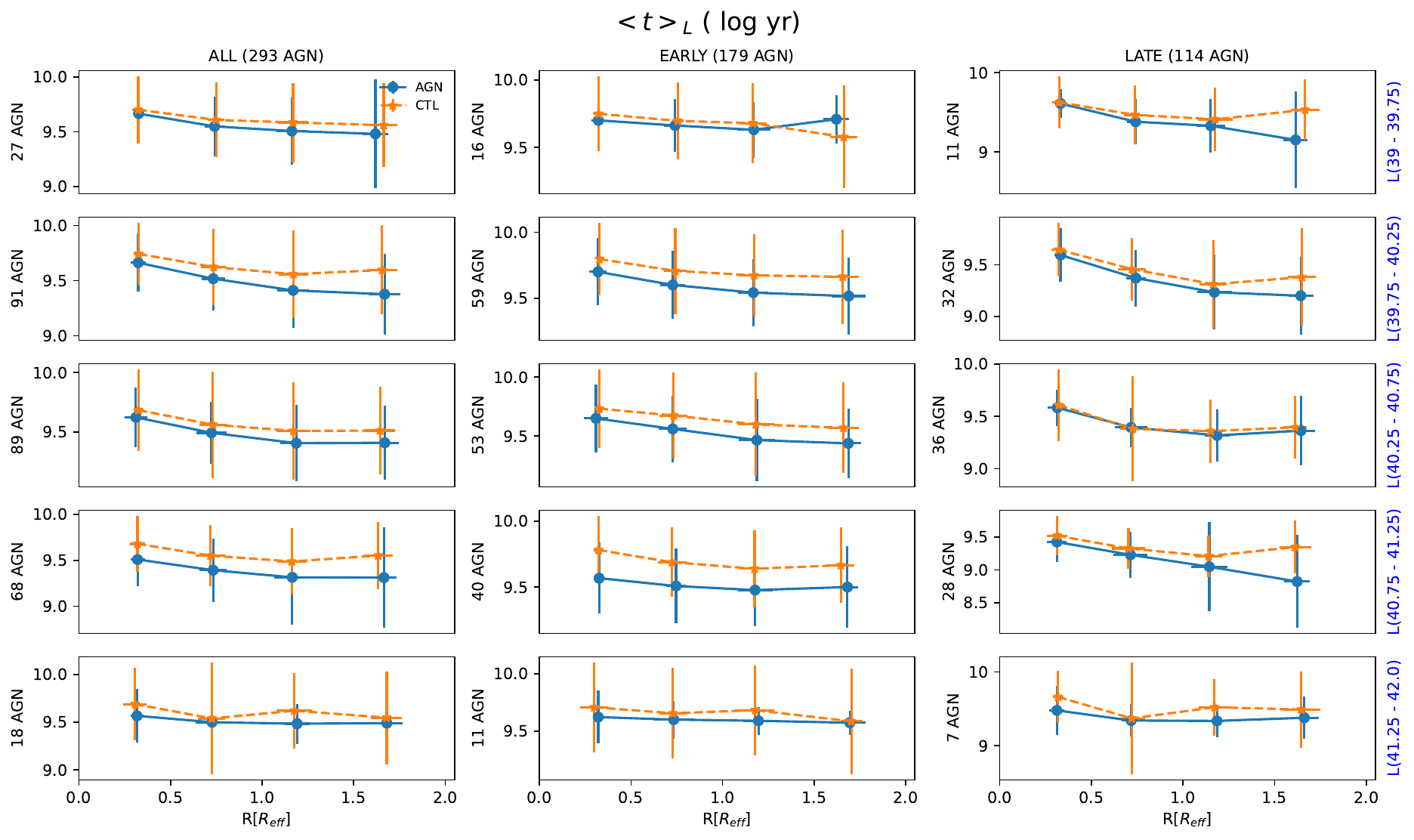}
\caption{Light-weighted mean age {\it versus} radius. The labels are the same as in Fig.~\ref{xyy}.} 
 \label{Mage}
 \end{figure*}

The quantitative comparison of individual sources is very difficult, and therefore we show the general behaviour of the stellar population parameters along the galaxies and the comparison between AGN and controls. We show their variations with radii and with [\ion{O}{iii}] luminosity from Fig.~\ref{xyy} to Fig.~\ref{Mage}. The L([\ion{O}{iii}]) bins where taken from \citet{Rembold+17}.  We have also separated the sample into early- and late-type galaxies. The Hubble types were taken from \citet{Vazquez-Mata+22} and are listed in Tab.~\ref{tableagns}. We have classified the sources with the parameter {\sc T-Type} $\leq$ 2 as early-type and with  2$<$ {\sc T-Type} $<$ 10 as late-type, while objects with  {\sc T-Type} $\geq$ 10 were classified as Irregular and are not used in our analysis.

What clearly emerges from this choice is that for the younger ages ($t<$56\,Myr) no differences are detected when comparing AGNs hosts and control galaxies. However, when focusing on R~$<$~0.5\,R$_e$ the values for this component seem to be biased towards a slight increase on the mean values for AGN hosts with higher [\ion{O}{iii}] luminosity, when compared to their corresponding control galaxy counterparts. In addition, this component's contribution is generally low ($\lesssim $ 10 percent) with the possible difference between AGNs and controls being around 2 percent and within the uncertainties. For the less luminous AGNs a positive  gradient is seen, while a slightly negative gradient is seen for the higher luminosity ones. This suggests that the more luminous AGNs may be facing a rejuvenation when compared with the less luminous ones. However, we call attention to the fact that a reddened young SP can mimic an AGN featureless continuum component \citep[see][for more details]{CidFernandes+04,CidFernandes+05,Riffel+09,Riffel+22}, thus, this young component has to be analyzed together whit the FC. In Fig.~\ref{fc} we show the behavior of the FC component as a function of R and L([\ion{O}{iii}]). In general, the values are very small ($\lesssim$5 percent), and slightly increase outwards\footnote{We attribute the increase in the value of FC with R, for both AGNs and controls, as due to an increase of the noise in the spaxels, such that the absorption features are washed away and the FC component becomes hardly distinguishable from the underlying stellar continuum \citep[see][and references]{CidFernandes+14,CidFernandes+18}.}. For the two higher luminosity bins a higher value for FC  (of $\sim$10 percent) is seen for R$<$0.5\,R$_e$ , and it becomes quite different than the contributions found for the control sample ($\sim$2 percent), which we interpret as due to a true AGN FC contribution, while for the remaining luminosity bins and for larger radii we interpret it as due to a reddened young stellar population.

From Fig.~\ref{xi} it is clear that there is a significant difference for the intermediate age (100~Myr $<t\leq$ 2~Gyr) population between AGN hosts and control galaxies. This difference increases with luminosity and is seen for both early and late-type galaxies. Also, a positive gradient is seen with the contribution of the intermediate population increasing outwards. 

When looking at the old component ($t \ge$ 5\,Gyr), a clear difference between the AGN hosts and their controls is seen, becoming more evident for higher luminosities. A slightly steeper gradient, with the fraction of the old population decreasing with R, is seen in the case of AGN hosts when compared with controls. 

The above results can be summarised when we looking how the light-weighed mean age ($<t>_L$, Fig.~\ref{Mage}) changes with R. What becomes clear when looking for  R$\lesssim$0.5\,R$_e$, is that AGN hosts become younger as the AGN luminosity increases (for both, Early and Late-type sources), but, the mean ages are smaller for AGN hosts over R values up to 1.75\,R$_e$, indicating that the stellar population of the host galaxies is affected by the nuclear activity. In addition, a negative gradient of $<t>_L$ with  radius is also found, with a flattening for higher AGN [\ion{O}{iii}] luminosities.

As can be seen from Figs.\ref{xyy} to \ref{Mage}, in general, we observe that the fraction of young and intermediate-age stellar populations increases with the radius, while in the case of the old population, it decreases. These findings are in agreement with previous results \citep[e.g.][and references therein]{Sanchez+13,Ibarra-Medel+16,Goddard+17,Mallmann+18,Sanchez+21}. Our results are similar to the results we have obtained for the first 62 AGNs observed by MaNGA \citep[][]{Mallmann+18}. When looking at controls and AGNs, separated between early- and late-type sources, the sense of our gradients are opposed to those found by \citet{Goddard+17}. For early-types we derive negative gradients while theirs are slightly positive.  The results presented here are in full agreement with those by \citet{Ibarra-Medel+16}, who found that the radial stellar mass growth histories of early-type galaxies increase outward, though with a trend much less pronounced than that of their late-type galaxies. The decreasing gradient we observed for the old component suggests that the relative contribution of the more recent star formation increases going outward.  Both, early and late-type sources present a significant contribution of the intermediate ($\sim$20/40 percent for Early/Late, respectively, for R$<$0.5\,R$_e$, in light fractions) and young populations ($\sim$5/7 percent for Early/Late, respectively, for R$<$0.5\,R$_e$, in light fractions). This indicates that even in early-type sources residual star-formation is taking place, suggesting that the galaxies are facing a rejuvenation process \citep[e.g.][]{Mallmann+18,Martin-Navarro+22} or still forming stars with the gas that they have still available \citep[e.g. they have not finished their SF over cosmic time, see][and references]{Eales+17,Dimauro+22,Paspaliaris+23}, thus they are in fact (predominantly) red and {\it not } dead sources \citep[][]{Salvador-Rusinol+20,Salvador-Rusinol+21,deLorenzo-Caceres+20,Benedetti+23}.


\begin{figure*}
 	\includegraphics[width=2.\columnwidth]{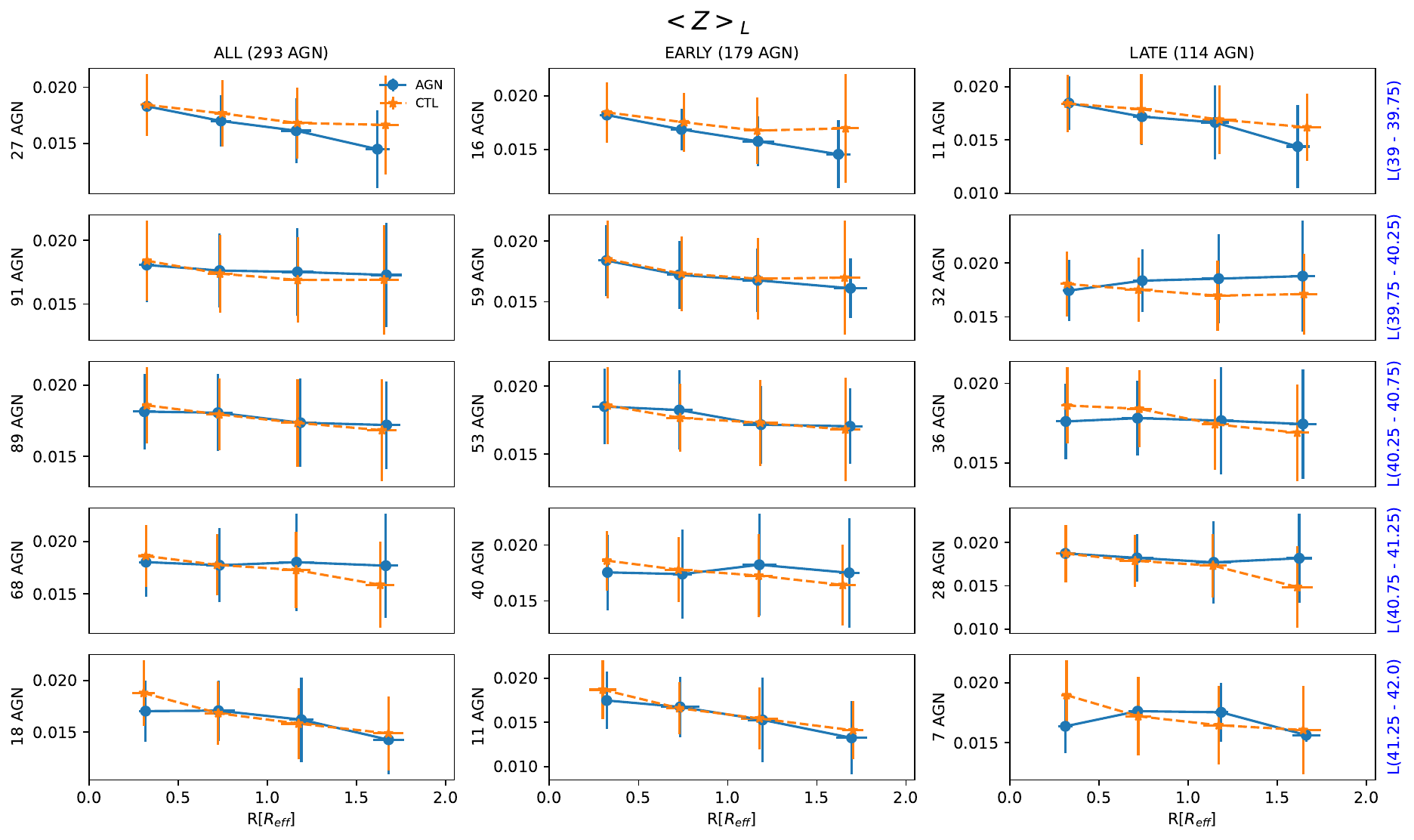}
\caption{Light-weighted mean metallicity {\it versus} radius. The labels are the same as in Fig.~\ref{xyy}.}  \label{MZ}
 \end{figure*}

\subsubsection{Dependence on the AGN luminosity}

In the case of the most luminous AGNs [log(L[\ion{O}{iii}]:41.25 - 42.00)] the young population contribution seems to rise to higher values for lower radii values. We interpret this as the fact that these AGNs are receiving an extra supply of gas and thus are more luminous. In fact, as shown in \citet{Riffel+22} using high angular resolution observations, found that the inner region of the AGNs is dominated by an intermediate age population ($<t>_L \lesssim $ 1.5 Gyr) and that a correlation between the bolometric luminosity of AGNs with the mean age of their stellar populations is observed in the sense that the more luminous AGNs have larger amounts of intermediate-age stars.  This correlation was interpreted as due to a delay between the formation of new stars and the triggering/feeding of the AGN. These intermediate-age stars do provide (via mass loss through stellar evolution) an extra amount of gas that will reach the supermassive black hole (SMBH). Such gas has a low velocity (a few hundred \kms) and is accreted together with the gas that is already flowing towards the central region of the host galaxy \citep[e.g.][]{Cuadra+06,Hopkins+12,Storchi-Bergmann+19}. In fact, young to intermediate-age populations are detected in the inner region of galaxies \citep[e.g.][]{Dottori+05,Davies+05,Riffel+07,Riffel+15,Salvador-Rusinol+20,Salvador-Rusinol+21,deLorenzo-Caceres+20}. Such population is dominated by short-lived stars ($t\simeq $0.2 -- 2\,Gyr; $\rm M\simeq 2 - 6\,M_{\odot}$) who do eject a significant amount of material to the nuclear environment. This recycled material can cool down forming new stars \citep[e.g.][]{Salvador-Rusinol+20,Salvador-Rusinol+21,deLorenzo-Caceres+20,Benedetti+23} and/or fuel the SMBH with an extra amount of gas, making the AGN brighter or triggering it. This would explain the fact that the most luminous AGNs in our sample do show larger fractions of this intermediate-age population when compared with their control sources.

\subsubsection{The stellar Metallicity}

In Fig.~\ref{MZ} we show the mean stellar metalicity. Overall, the trend for both the AGN hosts and control galaxies shows a decreasing gradient with increasing radius. A similar behaviour is found by \citet{Goddard+17} and \citet{Sanchez+20}, who found a negative gradient for the stellar metallicity for high mass galaxies and positive for low mass ones. For our sample, when all sources are taken together (e.g. left side pannel) the most luminous AGN hosts show a slightly shallower gradient when compared with the controls\footnote{This is in full agreement with the findings of \citet{Sanchez+20}, since the AGN luminosity can be taken as a proxy for the galaxy mass (thus the control galaxies for the most luminous AGNs would be more massive).}. Additionally, for regions R~$<$~0.5\,R$_e$ the AGN hosts metallicity is biased towards lower values when compared with the control galaxies and larger radii values. In fact, this is in agreement with the finding of \citet{doNascimento+22} who derive the gas metallicity (traced by the O/H abundance) of 108 Seyfert galaxies from MaNGA, and found that the inner regions of these galaxies display lower abundances than their outer regions. Additionally, \citet{Armah+23} have shown that the more luminous Seyfert galaxies have lower gas metalicities. However, these results can not be straightforward compared since the gas metallicity is a recent picture of the chemical evolution of the galaxies, while the mean stellar metalicity is a mean over the galaxies' lifetime (e.g. the mean over all the stellar generations). Nevertheless, our stellar metalicity estimates are in agreement with those obtained through the gas phase abundances. \rev{}

\subsection{Spatially resolved diagnostic diagrams: BPT and WHAN}

It is known that AGN produces a much harder radiation field than main sequence stars. A widely used tool to identify and classify emission-line sources according to their dominant radiation field are the diagrams proposed by \citet[][hereafter BPT]{Baldwin+81}  which are based on line ratios between high and low ionization potential species (e.g. [\ion{N}{ii}]/\ha\ $\times$ [\ion{O}{iii}]/\hb). However, the BPT diagrams cannot discriminate between genuine
low-ionization AGN and emission-line galaxies whose ionizing photons are produced in the atmospheres of  hot low-mass evolved stars (HOLMES). To overcome this \citet{CidFernandes+10} have proposed a diagram that uses as a discriminator the equivalent width (EW) of the \ha\ emission line. They have shown that galaxies with EW(H$\rm _\alpha) < $~3\AA, may not be ionized by an AGN but by the HOLMES. This diagram is known as the WHAN diagram since it involves the [\ion{N}{ii}]/\ha\ and EW(\ha). 

\begin{figure*}
	\includegraphics[scale=0.58]{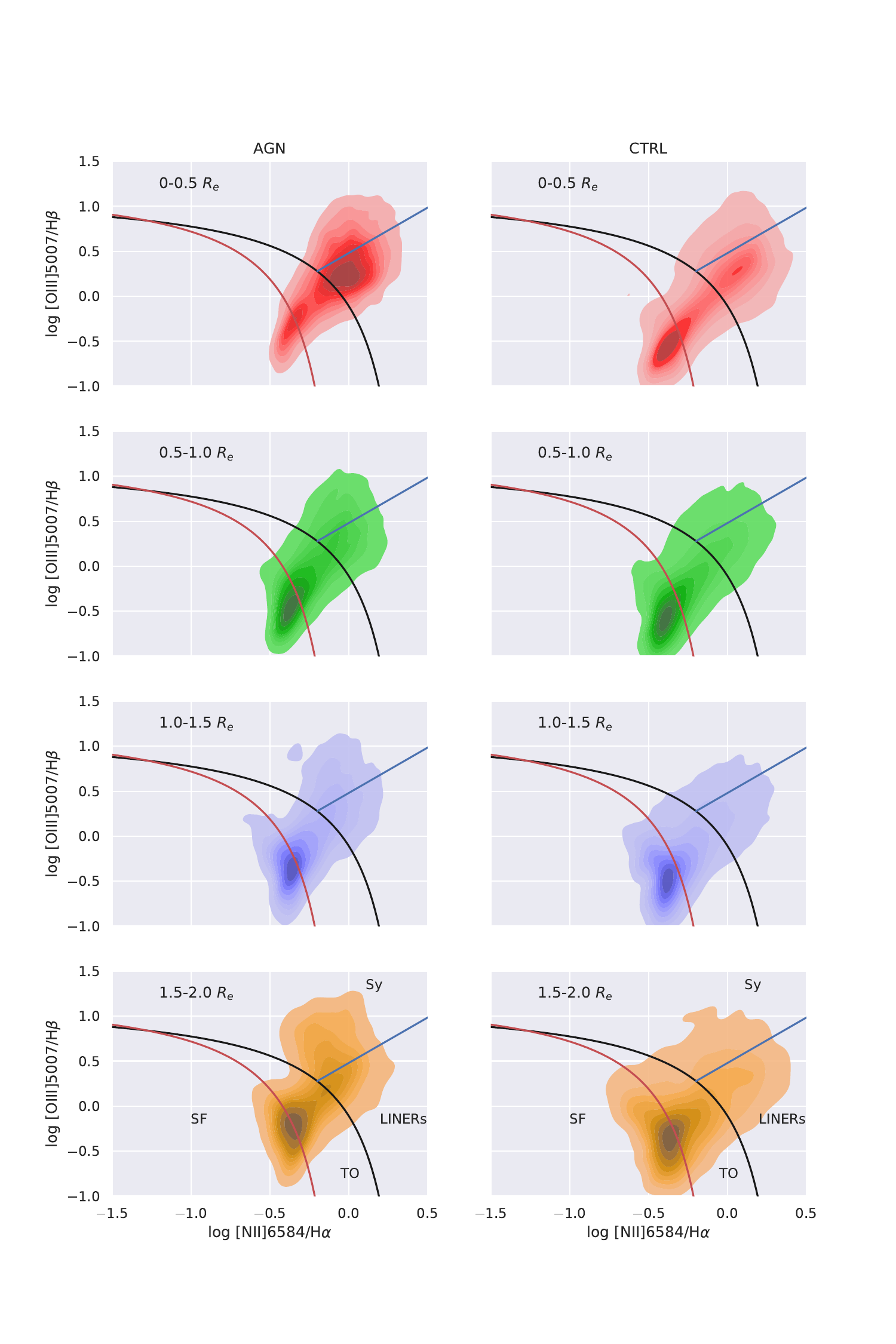}
    \caption{BPT diagnostic diagram for our AGN and control samples. In these diagrams, we show all the spaxels where SNR in the continuum is $\geq 10 $ and the [\ion{O}{iii}]$\lambda$ 5007\AA\ emission line is detected with a 3$\sigma$ confidence level. Different radial values are shown, from top to bottom: 0 -- 0.5\,$R_e$ (red), 0.5 -- 1.0\,$R_e$ (green), 1.0 -- 1.5\,$R_e$ (blue), and 1.5 -- 2.0\,$R_e$ (yellow). Black and red lines correspond to the empirical and theoretical criteria to separate AGN-like and H ii-like objects proposed by \citet{Kewley+01} and \citet{Kauffmann+03b}, respectively. The blue line represents the separation between AGN-like and LINERS-like sources proposed by \citet{CidFernandes+10}.
    	}
    \label{bpt}
\end{figure*}

\begin{figure*}
	\includegraphics[scale=0.58]{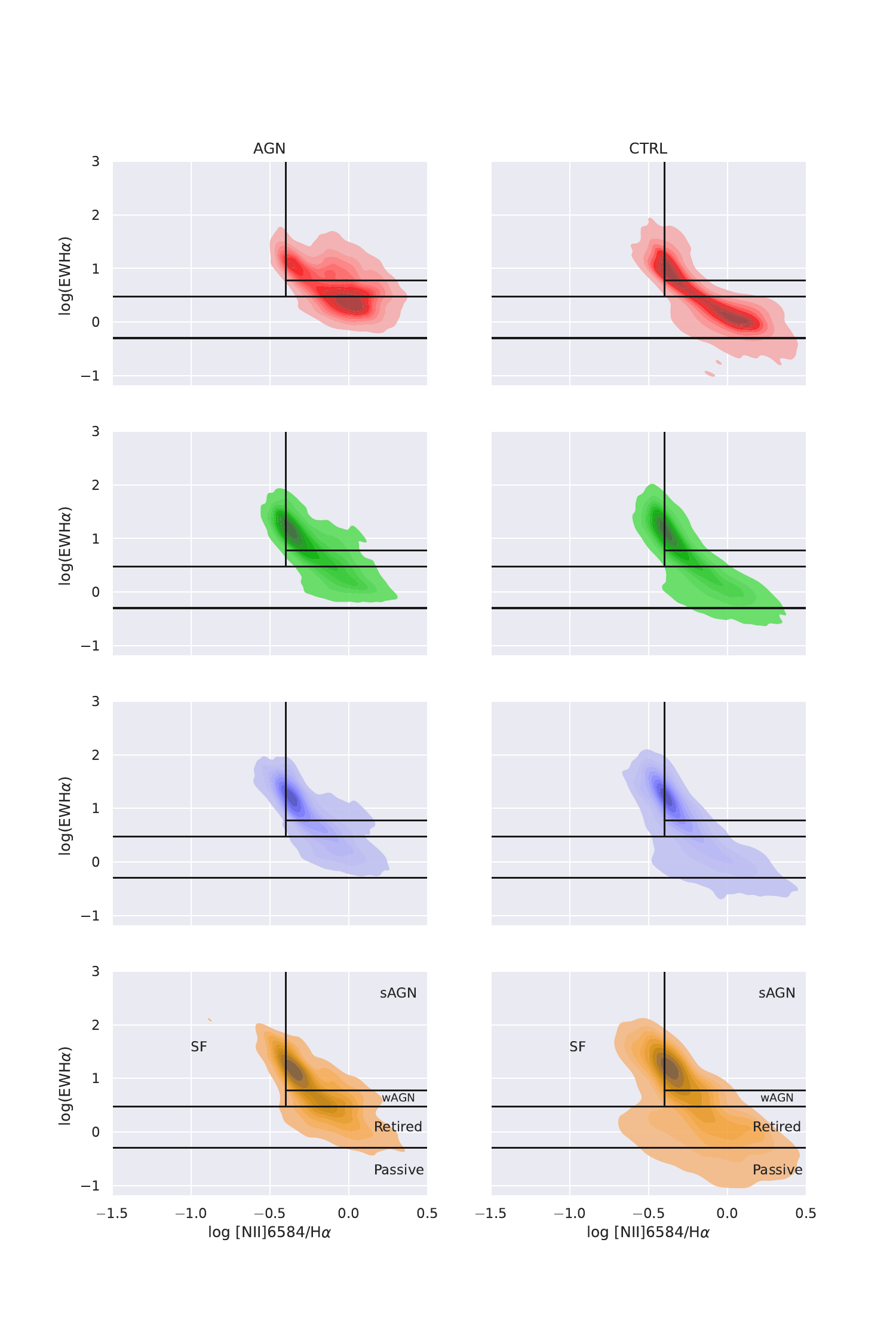}
    \caption{ {\sc WHAN} diagnostic diagram for our AGN and control samples for the same data-points as in Fig.~\ref{bpt}
    	}
    \label{whan}
\end{figure*}

In Fig.~\ref{bpt} and Fig.~\ref{whan} we present the BPT and {\sc WHAN} diagrams, respectively for AGNs and controls. These diagrams have been built with all the spaxels where the SNR in the continuum is $\geq 10 $ and the [\ion{O}{iii}]$\lambda$ 5007\AA\ emission line is detected with a minimum of 3$\sigma$ confidence level, and for four different radial regions in the galaxies.

As can be seen in Fig.~\ref{bpt}, the BPT diagrams reveal that most of the AGN emission is concentrated  in the inner 0.5\,$R_e$ region of the galaxies (with Sy/LINER classification), as well as that the bulk of the emission for the control galaxies is located in the SF-dominated region. In addition, for regions outside the inner 0.5\,$R_e$, most of the AGN hosts and control galaxies become SF-dominated and present very similar distributions. In summary, when moving radially across the galaxy (increasing $R_e$) what is seen is that the ionization pattern moves from the AGN-dominated region to the SF-dominated one, crossing the transition region.

On the other hand, the same behavior is not clearly seen in the WHAN diagram (Fig.~\ref{whan}).
While the bulk of the spaxels in the inner 0.5\,$R_e$ of AGN hosts is located at consistently larger EW(\ha ) values ($\sim 0.5$\,dex) than the ``retired'' wing of the control galaxies, spaxels in both samples also distribute along a diagonal path crossing the wAGN/sAGN regions all the way to the SF region. This extended path in the WHAN diagram corresponds to the displacement from AGN/LI(N)ER to SF-ionization in the BPT diagram across the transition region.

It is important to note that the WHAN diagram seems effective at separating the ``forgotten" population of retired galaxies from true AGN-ionized ones; however, the transition region (e.g. with photons from HOLMES/SF+AGN) is not explicitly defined in this diagram. The distinction between AGN and SF objects is much more diffuse. At radial distances larger than 0.5\,$R_e$, spaxels tend to be located in, or close to, the SF region, both for AGN hosts and control galaxies, but a significant fraction of spaxels is still located in the sAGN/wAGN region. This may be due to the fact that, as we move toward higher radii values, we observe regions where the role played by SF on the gas excitation increases, while that of the AGN decreases \citep{Sanchez+20}, and so the EW(\ha) increases, making the locus of the region move towards the weak-AGN/SF dominated region. In the studies of  \citet[][]{Deconto-Machado+22} as in Gatto et al. ({\it in preparation}), on the same sample of AGN and control galaxies discussed here, we have found continuity in the kinematic properties of AGN and controls, in terms of the line widths of the [\ion{O}{iii}]$\lambda$5007\AA\ emission lines vs. L([\ion{O}{iii}]). This can be interpreted as due to the fact that even in the regions with  EW(\ha)~$< 3 \AA$ the gas may be still ionized by a mix of photons coming both from the AGN and from HOLMES. This becomes even more clear when BPT and WHAN are considered together.

\begin{figure*}
	\includegraphics[scale=0.58]{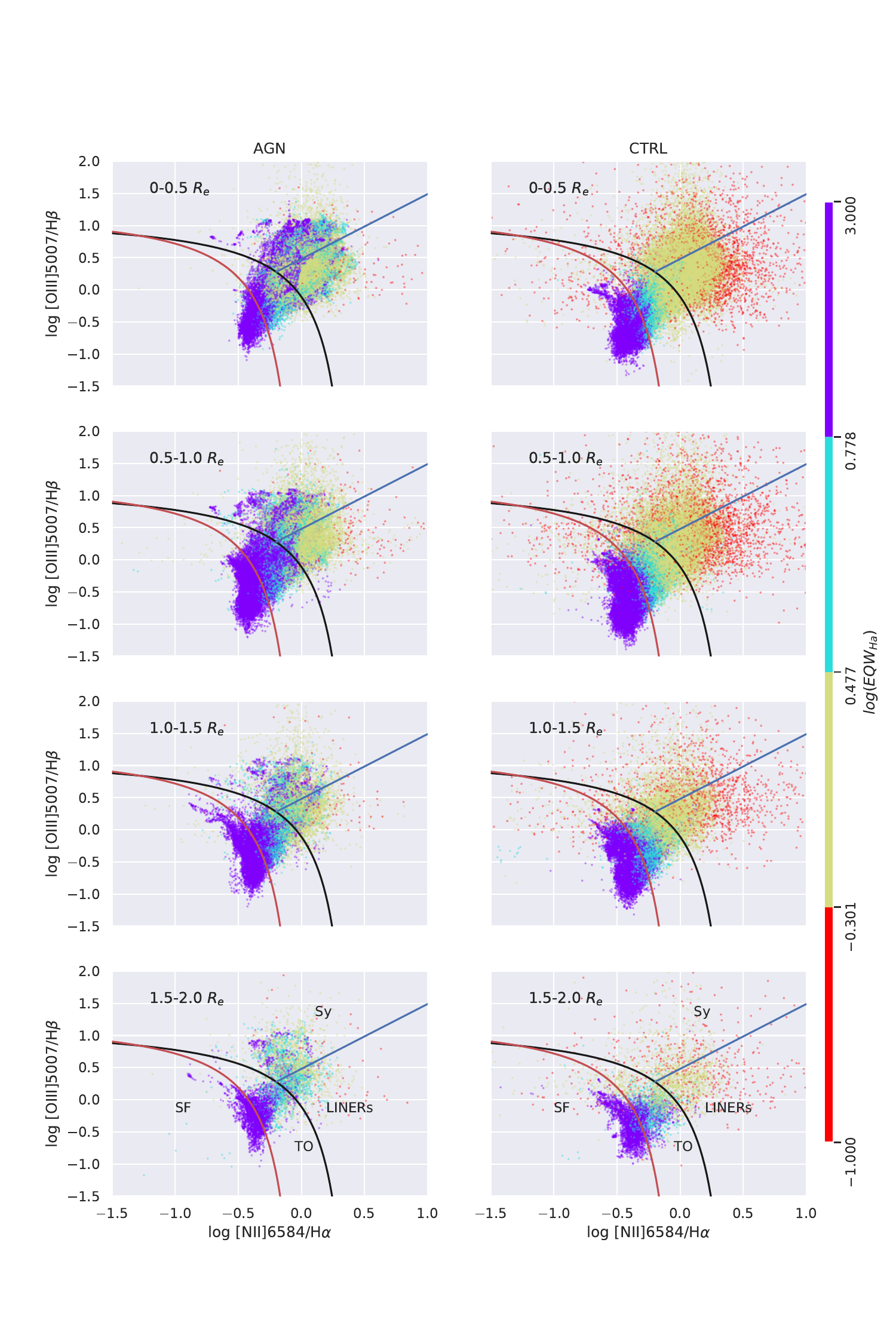}
    \caption{BPT diagnostic diagram for our AGN and control samples for distinct radial distances (from top to bottom: 0 -- 0.5\,$R_e$, 0.5 -- 1.0\,$R_e$, 1.0 -- 1.5\,$R_e$, and 1.5 -- 2.0\,$R_e$) with each spaxel colored according to the equivalent width (EW) of the \ha\ emission line (using the limits of the {\sc WHAN} for EW\ha). The lines are the same as in Fig.~\ref{bpt}.
    	}
    \label{bpt_scatter}
\end{figure*}

In order to try to clarify the behavior described above, we have considered the BPT and WHAN together in Fig.~\ref{bpt_scatter}: as can be seen in this figure the spaxels with larger EW(\ha) when moving to larger radii move to the SF dominated region, clearly indicating that to do a proper activity classification, besides the line ratios and  EW(\ha), aperture effects also need to be considered and that the bulk of the emission outside the central region is dominated by star-forming processes \citep{Sanchez+20}.  In fact, this aperture-dependent (mis)classification has been reported by \citet{Alban+23}. They found that the number of fake-AGN increases significantly for values larger than 1.0$R_e$ and they attribute it to increased contamination from diffuse ionized gas. \citet{Sanchez+20} shows that the ionization in the overall extension of galaxies is dominated by stellar processes (hot young stars or HOLMES), with the AGN emission being dominant in the central region of active galaxies.

\section{Final Remarks}\label{sec:final_rem}

We present here spaxel-by-spaxel stellar population fits for the $\sim$10 thousand MaNGA datacubes. We provide multiple extension fits files, nominated as {\sc megacubes}, with maps of light- and mass-fraction contribution of each stellar population component, light-weighted and mass-weighted mean age and mean metallicity, reddening, star formation rates via stellar population (over the last  1\,Myr, 5.6\,Myr, 10\,Myr, 14\,Myr, 20\,Myr, 32\,Myr, 56\,Myr, 100\,Myr, and 200\,Myr), binned population vectors in light- and mass-fraction in different age ranges (xyy$\leq$10\,Myr; 14\,Myr$\rm<xyo\leq$56\,Myr; 100\,Myr$\rm<xiy\leq$500\,Myr; 630\,Myr $\rm<xii\leq$800\,Myr; 890\,Myr$\rm<xio\leq$2.0\,Gyr;  5.0\,Gyr$\rm<xo\leq$13\,Gyr). Parameters characterising the emission-line profiles of H$\beta$, [O\,{\sc iii}]$\lambda\lambda$4959,5007, He\,{\sc i}$\lambda$5876, [O\,{\sc i}]$\lambda$6300, H$\alpha$ [N\,{\sc ii}]$\lambda\lambda$6548,6583 and [S\,{\sc ii}]$\lambda\lambda$6716,6731, are also provided for each spaxel. 
All the {\sc megacubes} are available through a web interface (\url{https://manga.linea.org.br/} or \url{https://www.if.ufrgs.br/~riffel/software/megacubes/}), where each one of them can be inspected, downloaded, and interacted with via different plots.

We analysed the {\sc megacubes} for the final AGN sample (293) and control galaxies (586) in the MaNGA survey selected according to the criteria from \citet{Rembold+17}, in which the control galaxies are matched to the AGN according to stellar mass, distance, inclination, and galaxy type.

We have also presented a global analysis of the stellar population of the final set of AGNs and compared them with the control galaxies sample. We found that the young and intermediate-age populations show an outwards increasing gradient for AGNs and controls, while the old component decreases outwards, indicating that the galaxies are facing a rejuvenation process. We find that the fraction of intermediate-age stellar population is higher in AGN hosts than that found in the control sample, and this difference becomes larger for higher-luminosity AGNs. This has been interpreted by us as the fact that an extra amount of gas is available in these more luminous sources and that most likely it originates from mass loss from the intermediate-age stellar population. We have also found that the mean metalicities show a similar trend for AGN hosts and control galaxies showing a decreasing gradient with increasing radius, with AGN hosts showing a slightly shallower gradient. Additionally, for regions R~$<$~0.5\,R$_e$ the AGN hosts metallicity is biased towards lower values when compared with the control galaxies, as well as when compared with larger radii values. This is in agreement with the behavior found when studying the gas phase abundances. 

Analysis of the global properties of the gas excitation via BPT and WHAN diagrams have also been presented.
The spatially resolved BPT diagrams reveal that the AGN emission is concentrated in the inner 0.5\,$R_e$ region of the galaxies (with Sy/LINER classification), as well as that the bulk of the distribution for the control galaxies is located in the SF-dominated region. Additionally, besides the dependence on aperture size for proper activity type classification, we show that the BPT diagram has to be used together with the strength of the \ha\ equivalent width. We thus present a ``composite" BPT+WHAN diagram to be used for as a more comprehensive diagnostic of the emitting gas excitation in galaxies.


\section*{Acknowledgments}
We thank the anonymous referee for useful suggestions which helped to improve the paper. 
RR acknowledges support from the Fundaci\'on Jes\'us Serra and the Instituto de Astrof{\'{i}}sica de Canarias under the Visiting Researcher Programme 2023-2025 agreed between both institutions. RR, also acknowledges support from the ACIISI, Consejer{\'{i}}a de Econom{\'{i}}a, Conocimiento y Empleo del Gobierno de Canarias and the European Regional Development Fund (ERDF) under grant with reference ProID2021010079, and the support through the RAVET project by the grant PID2019-107427GB-C32 from the Spanish Ministry of Science, Innovation and Universities MCIU. This work has also been supported through the IAC project TRACES, which is partially supported through the state budget and the regional budget of the Consejer{\'{i}}a de Econom{\'{i}}a, Industria, Comercio y Conocimiento of the Canary Islands Autonomous Community. RR also thanks to Conselho Nacional de Desenvolvimento Cient\'{i}fico e Tecnol\'ogico  ( CNPq, Proj. 311223/2020-6,  304927/2017-1, 400352/2016-8, and  404238/2021-1), Funda\c{c}\~ao de amparo \`{a} pesquisa do Rio Grande do Sul (FAPERGS, Proj. 16/2551-0000251-7 and 19/1750-2), Coordena\c{c}\~ao de Aperfei\c{c}oamento de Pessoal de N\'{i}vel Superior (CAPES, Proj. 0001).

RAR acknowledges support from CNPq and FAPERGS. GSI acknowledges support from CNPq  (142514/2018-7) and FAPESP (Fundação de Amparo à Pesquisa do Estado de São Paulo, Proj. 2022/11799-9). We are also grateful to  LIneA's IT team which has supported the creation of the web portal and the INCT e-Universo.
CRA acknowledges support from the projects “Feeding and feedback in active galaxies”, with reference PID2019-106027GB-C42, funded by MICINN-AEI/10.13039/501100011033, and ``Quantifying the impact of quasar feedback on galaxy evolution'', with reference EUR2020-112266, funded by MICINN-AEI/10.13039/501100011033 and the European Union NextGenerationEU/PRTR.

SDSS is managed by the Astrophysical Research Consortium for the Participating Institutions of the SDSS Collaboration including the Brazilian Participation Group, the Carnegie Institution for Science, Carnegie Mellon University, the Chilean Participation Group, the French Participation Group, Harvard-Smithsonian Center for Astrophysics, Instituto de Astrofisica de Canarias, The Johns Hopkins University, Kavli Institute for the Physics and Mathematics of the Universe (IPMU) / University of Tokyo, the Korean Participation Group, Lawrence Berkeley National Laboratory, Leibniz Institut f\"ur Astrophysik Potsdam (AIP), Max-Planck-Institut f\"ur Astronomie (MPIA Heidelberg), Max-Planck-Institut f\"ur Astrophysik (MPA Garching), Max-Planck-Institut f\"ur Extraterrestrische Physik (MPE), National Astronomical Observatories of China, New Mexico State University, New York University, University of Notre Dame, Observat\' orio Nacional / MCTI, The Ohio State University, Pennsylvania State University, Shanghai Astronomical Observatory, United Kingdom Participation Group, Universidad Nacional Aut\'onoma de M\'exico, University of Arizona, University of Colorado Boulder, University of Oxford, University of Portsmouth, University of Utah, University of Virginia, University of Washington, University of Wisconsin, Vanderbilt University, and Yale University.

\section*{Data Availability}

 The data underlying this article are available under SDSS collaboration rules, and the by-products are available at: \href{https://manga.linea.org.br/}{https://manga.linea.org.br/} or \url{https://www.if.ufrgs.br/~riffel/software/megacubes/} following the SDSS collaboration rules.

\bibliographystyle{mnras}
\bibliography{RefsZotero.bib}



\includepdf[pages=-]{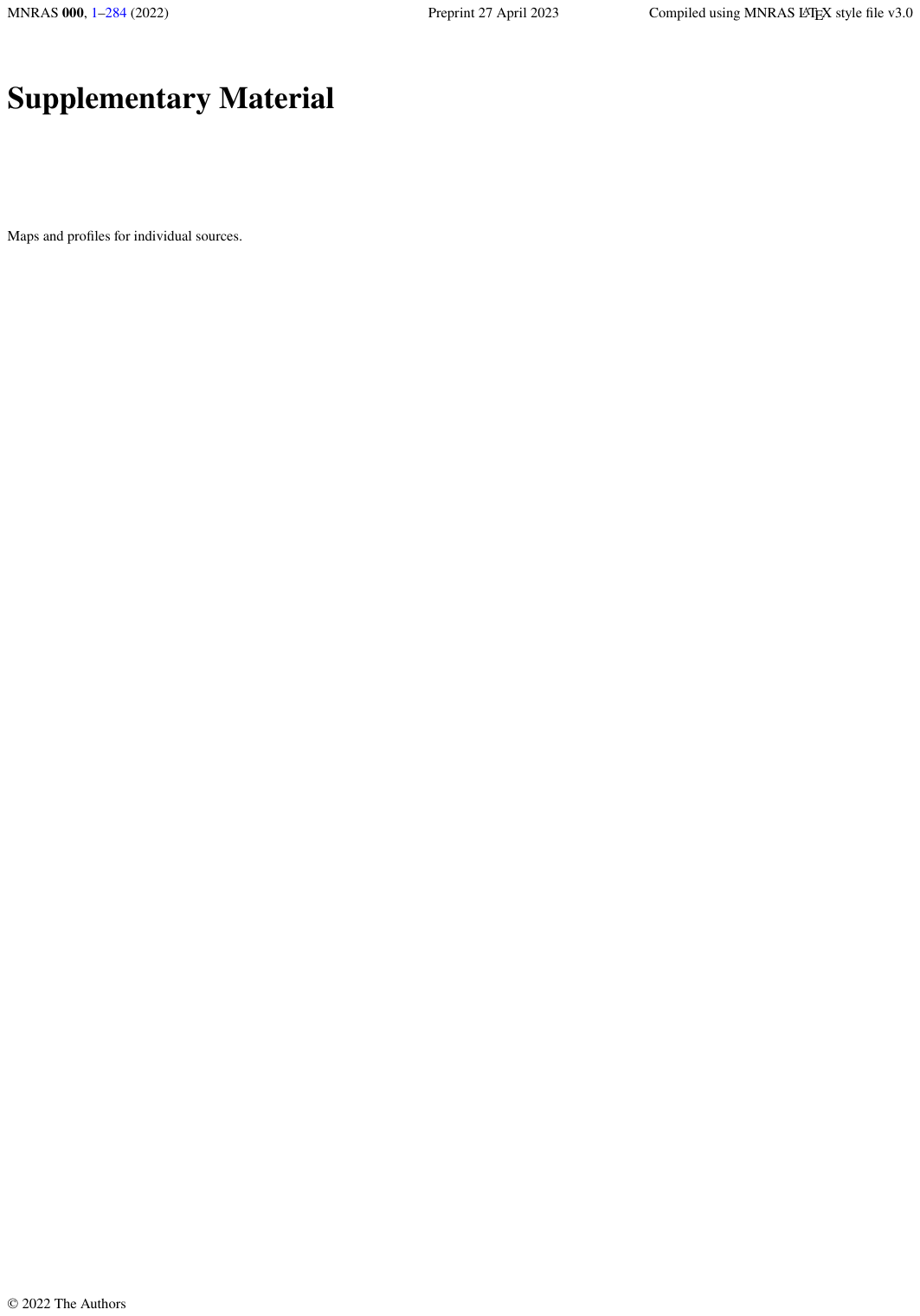}

\end{document}